\documentclass[12pt]{article}

\usepackage{amsmath,amssymb,cite}
\usepackage{graphicx,color}

\newcommand{\f}[2]{\frac{#1}{#2}}
\newcommand{\ko}[1]{\left( #1 \right)}
\newcommand{\kko}[1]{\left[ #1 \right]}

\newcommand{\bmt}[1]{{{\mbox{\boldmath$ #1 $}}}}
\newcommand{\komoji}[1]{\mbox{$#1$}}
\newcommand{\ord}[1]{{\mathcal O\mbox{\small $\left(#1\right)$}}}

\renewcommand{\thefootnote}{\fnsymbol{footnote}}

\DeclareMathOperator{\tr}{Tr}

\def\ds{\displaystyle}
\def\pa{\partial}
\def\eq{\equiv}
\def\be{\beta}
\def\al{\alpha}

\def\th{\theta}
\def\Th{\Theta}
\def\no{\nonumber}
\def\lam{\lambda}

\def\ep{{\epsilon}}

\def\check{{\mbox{\small $\surd$}}}
\def\batsu{{\mbox{\small $\times$}}}

\def\bul{{\small ~\,$\bullet$~\,}}

\numberwithin{equation}{section}

\begin{document}

\quad
\vspace{-2.0cm}

\begin{flushright}
{\bf December 2007}\\
\vspace{0.3cm}
{\small DAMTP\,-\,2007\,-\,126 \hfill \\
UT\,-\,07\,-\,20 \hfill \\}
\end{flushright}


\begin{center}
\Large\bf 

Singularities of the Magnon Boundstate S-Matrix

\end{center}

\vspace*{1.0cm}
\centerline{
{\sc Nick Dorey$^{\dagger,\, a}$
~and~
Keisuke Okamura$^{\ddagger,\, b}$}}
\vspace*{0.5cm}
\begin{center}
${}^{\dagger}$\emph{DAMTP, Centre for Mathematical Sciences, Cambridge 
University,\\
Wilberforce Road, Cambridge CB3 OWA, UK} \\
\vspace{0.3cm}
\vspace{0.3cm}
${}^{\ddagger}$\emph{Department of Physics, Faculty of Science, 
University of Tokyo,\\
Bunkyo-ku, Tokyo 113-0033, Japan} \\
\vspace*{0.6cm}
${}^{a}$\,{\tt N.Dorey@damtp.cam.ac.uk}\qquad 
${}^{b}$\,{\tt okamura@hep-th.phys.s.u-tokyo.ac.jp}
\end{center}
\vspace*{1.0cm}

\centerline{\bf Abstract} 
\vspace*{0.5cm}

We study the conjectured exact S-matrix for the scattering of 
BPS magnon boundstates in the spin-chain description of planar 
${\cal N}=4$ SUSY Yang-Mills. The conjectured S-matrix exhibits both
simple and double poles at complex momenta. Some of these poles lie
parametrically close to the real axis in momentum space on the branch
where particle energies are positive. We show that
all such poles are precisely accounted for by physical processes involving one
or more on-shell intermediate particles belonging to the known BPS
spectrum.

\vspace*{1.0cm}
\vfill

\thispagestyle{empty}
\setcounter{page}{0}
\setcounter{footnote}{0}
\setcounter{figure}{0}
\renewcommand{\thefootnote}{\arabic{footnote}}
\newpage

\section{Introduction and Preliminaries}

The correspondence between singularities of the S-matrix and on-shell
intermediate states is a standard feature of quantum field
theory. It can be understood as a consequence of the 
analyticity and unitarity of the S-matrix\footnote{More precisely, only those
singularities in a suitably defined ``physical region'' need have an
explanation in terms of on-shell states.}. In a recent paper \cite{Dorey:2007xn}, this
correspondence was investigated in the context of the spin-chain
description of planar $SU(N)$ ${\cal N}=4$ SUSY Yang-Mills \cite{Minahan:2002ve, Beisert:2003tq, Beisert:2003yb}. 
In particular, the 
poles of the conjectured exact S-matrix for magnon scattering 
\cite{Staudacher:2004tk, Beisert:2005tm, Beisert:2006ib, Beisert:2006ez} 
were precisely accounted for by considering processes involving the
exchange of one or more BPS magnon boundstates. The goal of the
present paper is to extend this investigation to the corresponding
S-matrix for the scattering of the boundstates themselves.

Initially we will focus on the $SU(2)$ sector of the ${\cal N}=4$
theory. Operators in this sector are composed of two of the three
complex scalar fields of the theory, which we denote as $Z$ and $W$. 
Single trace operators have the form $\mathcal
O=\tr\ko{\Phi_{i_{1}}\Phi_{i_{2}}\dots}+{\rm permutations}$ with each
$\Phi$ being 
either $Z$ or $W$\,.
In particular, the BPS operator $\tr\ko{ZZ\dots}$ made up of only $Z$s
is the ferromagnetic ground state 
for the SYM spin-chain.

The asymptotic Bethe ansatz equation \cite{Beisert:2005fw} for the $SU(2)$
sector of the theory reads, 
\begin{equation}
e^{i p_{j}}=\prod_{k\neq j}^{M} s(p_{j},p_{k})
\qquad \mbox{for}\quad 
j=1,\dots, M\,,
\end{equation}
where $M$ is the number of magnons, and the S-matrix \cite{Staudacher:2004tk} 
is given by
\begin{equation}
s(p_{j},p_{k})
=S_{\rm BDS}^{-1}(p_{j},p_{k})\cdot
\sigma(p_{j}, p_{k})^{-2}\,,\qquad 
S_{\rm BDS}^{-1}(p_{j},p_{k})=\f{u(p_{j})-u(p_{k})-(i/g)}{u(p_{j})-u(p_{k})+(i/g)}\,.
\label{S-matrix}
\end{equation}
and the rapidity function $u$ is given in 
terms of the momentum by
\begin{equation}
u(p)=\f{1}{2g}\cot\ko{\f{p}{2}}\sqrt{1+16g^{2}\sin^{2}\ko{\f{p}{2}}}\,.
\label{rap}
\end{equation}
Here $g$ is a gauge coupling constant related to the 't Hooft coupling $\lam=g_{\rm YM}^{2}N$ as $g\eq \sqrt{\lam}/4\pi$\,.
It is also convenient to introduce complex variables $x^{\pm}$ called
spectral parameters, which are related to the rapidity (\ref{rap}) via the formulae 
\begin{equation}
x^{\pm}(u)=x\ko{u\pm \mbox{\large $\f{i}{2g}$}}
\quad \mbox{where}\quad 
x(u)=\f{1}{2}\ko{u+\sqrt{u^{2}-4}}\,.
\label{x(u)}
\end{equation}

The factor $\sigma^{-2}$ appearing in (\ref{S-matrix}) is known as the
dressing factor \cite{Arutyunov:2004vx}. 
The conjectured exact expression for this function \cite{Beisert:2006ez} is
conveniently given as \cite{Dorey:2007xn},  
\begin{equation}
\sigma(x_{j}^{\pm}, x_{k}^{\pm})^{-2}=\ko{\f{R(x_{j}^{+},
    x_{k}^{+})R(x_{j}^{-}, x_{k}^{-})}
{R(x_{j}^{+}, x_{k}^{-})R(x_{j}^{-}, x_{k}^{+})}}^{-2}\,,\qquad 
R(x_{j}, x_{k})=e^{i\kko{\chi(x_{j}, x_{k})-\chi(x_{k}, x_{j})}}\,,
\label{BEHLS}
\end{equation}
where 
\begin{equation}
\chi(x_{j},x_{k})=-i\oint_{\mathcal C}\f{dz_{1}}{2\pi}\oint_{\mathcal
  C}\f{dz_{2}}{2\pi}
\f{\log\Gamma\ko{1+ig\ko{z_{1}+\f{1}{z_{1}}-z_{2}-\f{1}{z_{2}}}}}
{(z_{1}-x_{j})(z_{2}-x_{k})}\,.
\label{chi}
\end{equation}
The contours in (\ref{chi}) are unit circles $|z_{1}|=|z_{2}|=1$. 

The other factor $S_{\rm BDS}$ originates from the all-loop
Bethe ansatz proposed in \cite{Beisert:2004hm}. We will refer to it as
the BDS S-matrix in the following. For each solution of the Bethe
ansatz equations, the energy of the corresponding state is simply the
sum of the energies of the individual magnons. The energy of each
magnon is determined by the BPS dispersion relation
\cite{Beisert:2004hm, Beisert:2005tm} 
\begin{equation}
\epsilon_{j}= \sqrt{1+16g^{2}\sin^{2}\ko{\f{p_{j}}{2}}}\,.
\label{disp}
\end{equation} 
In terms of the spectral parameters, the magnon momenta and energies
are expressed as, 
\begin{align}
p_{j}&=p(x_{j}^{\pm})=\f{1}{i}
\log\bigg(\f{x_{j}^{+}}{x_{j}^{-}}\bigg)\,,\label{p}\\
\ep_{j}&=\ep(x_{j}^{\pm})=\f{g}{i}\kko{\bigg(x_{j}^{+}-\f{1}{x_{j}^{+}}\bigg)-\bigg(x_{j}^{-}-\f{1}{x_{j}^{-}}\bigg)}\,.
\label{Delta}
\end{align}
The dispersion relation (\ref{disp}) is equivalent to the constraint
\begin{equation}
\bigg(x_{j}^{+}+\frac{1}{x_{j}^{+}}\bigg) - 
\bigg(x_{j}^{-}+\frac{1}{x_{j}^{-}}\bigg)=\frac{i}{g}\,.
\end{equation}

\section{Boundstates and their S-matrix\label{sec:S-matrix}}

In term of the spectral parameters, BDS piece of the S-matrix takes the
form,
\begin{equation}
S_{\rm BDS}(x_{j}^{\pm}, x_{k}^{\pm})=\f{x_{j}^{+}-x_{k}^{-}}{x_{j}^{-}-x_{k}^{+}}\cdot
\f{1-1/(x_{j}^{+}x_{k}^{-})}{1-1/(x_{j}^{-}x_{k}^{+})}\,.
\label{S in x}
\end{equation}
We note the presence of a simple pole at $x_{j}^{-}=x_{k}^{+}$\,. 
As explained in \cite{Dorey:2006dq}, this pole indicates the formation
of a normalisable BPS boundstate of two magnons. In fact the theory
also contains a $Q$-magnon boundstate for each value of $Q>1$, 
related to a corresponding pole of the multi-particle S-matrix which can
be expressed as the product of two body factors by virtue of
integrability. These states were studied in detail in 
\cite{Dorey:2006dq, Chen:2006gq, Chen:2006gp}.
The spectral parameters of the constituent magnons in a $Q$-magnon
boundstate are,    
\begin{equation}
x_{j}^{-} = x_{j+1}^{+}
\qquad \mbox{for}\quad 
j=\komoji{1,\dots, Q-1}\,.
\end{equation}
The resulting boundstate of rapidity $U$ is described by introducing spectral
parameters 
\begin{equation}
X^{\pm}(U;Q)=x\ko{U\pm \mbox{\large $\f{iQ}{2g}$}}\,,\qquad 
\mbox{\em i.e.},\quad 
X^{+}\equiv x_{1}^{+}\,,\quad 
X^{-}\equiv x_{Q}^{-}\,.
\end{equation}
The total momentum $P$ and $U(1)$ charge $Q$ of the state are then expressed as
\begin{align}
P(X^{\pm})&=\f{1}{i}\log\ko{\f{X^{+}}{X^{-}}}\,,\\
Q(X^{\pm})&=\f{g}{i}\kko{\ko{X^{+}+\f{1}{X^{+}}}-\ko{X^{-}+\f{1}{X^{-}}}}\,.
\end{align}
One can also show the rapidity $U$ and energy
$E=\sum_{k=1}^{Q}\ep_{k}$ for the boundstate are related to the 
spectral parameters $X^{\pm}$ through the expressions
\begin{alignat}{3}
U(X^{\pm})	&=\f{1}{2}\kko{\ko{X^{+}+\f{1}{X^{+}}}+\ko{X^{-}+\f{1}{X^{-}}}}\,,\\
E(X^{\pm})	&=\f{g}{i}\kko{\ko{X^{+}-\f{1}{X^{+}}}-\ko{X^{-}-\f{1}{X^{-}}}}\,,
	\label{E}
\end{alignat}
or in terms of $P$ and $Q$\,,
\begin{alignat}{3}
U(P;Q)	&=\f{1}{2g}\cot\ko{\f{P}{2}}\sqrt{Q^{2}+16g^{2}\sin^{2}\ko{\f{P}{2}}}\,,\\
E(P;Q)	&=\sqrt{Q^{2}+16g^{2}\sin^{2}\ko{\f{P}{2}}}\,.\label{E2}
\end{alignat}
It is useful to note the following properties of those functions of $X^{\pm}$\,.
\begin{enumerate}
\item By an interchange $X^{+}\leftrightarrow X^{-}$\,, $P$\,, $Q$\,, $E$ change signs and only $U$ remains the same:
\begin{equation}
    \begin{aligned}
	U(X^{\pm})&=U(X^{\mp})\,,\quad &
	P(X^{\pm})&=-P(X^{\mp})\,,\cr
	Q(X^{\pm})&=-Q(X^{\mp})\,,\quad &
	E(X^{\pm})&=-E(X^{\mp})\,.
    \end{aligned}
    \label{interchange}
\end{equation}
\item By an inversion $X^{\pm}\leftrightarrow 1/X^{\pm}$ known as
  crossing transformation \cite{Janik:2006dc, Arutyunov:2006iu}, 
$P$ and $E$ change signs, while $U$ and $Q$ remain the same\,:
\begin{equation}
    \begin{aligned}
	U(X^{\pm})&=U(1/X^{\pm})\,,\quad &
	P(X^{\pm})&=-P(1/X^{\pm})\,,\cr
	Q(X^{\pm})&=Q(1/X^{\pm})\,,\quad &
	E(X^{\pm})&=-E(1/X^{\pm})\,.
    \end{aligned}
    \label{crossing}
\end{equation}
\end{enumerate}
Note also the spectral parameters for the boundstates can be written as
\begin{equation}
X^{\pm}(P;Q)=R(P;Q)\,e^{\pm iP/2}
\quad \mbox{with}\quad R(P;Q)=\f{Q+\sqrt{Q^{2}+16g^{2}\sin^{2}\ko{P/2}}}{4g\sin\ko{P/2}}\,.
\label{radius}
\end{equation}

\paragraph{}
Starting with the S-matrix (\ref{S-matrix}) for elementary magnons, it is
straightforward to obtain the corresponding S-matrix for 
magnon boundstates of arbitrary charges 
by fusion, as worked out in \cite{Chen:2006gq,Roiban:2006gs}. Because of factorisation of the multi-particle
S-matrix, the boundstate S-matrix is nothing other than the product of
two-body S-matrices describing all possible pair-wise scatterings
between the consitituent magnons.  
The procedure is illustrated schematically
in Figure 1.  
For two scattering boundstates (both in the same $SU(2)$ sector) with spectral parameters $Y_{1}^{\pm}$
and $Y_{2}^{\pm}$ and 
positive charges $Q_{1}\geq Q_{2}$\,, we find
\begin{align}
S(Y_{1}^{\pm},Y_{2}^{\pm})&=\prod_{j_{1}=1}^{Q_{1}}\prod_{j_{2}=1}^{Q_{2}}s(y_{j_{1}}^{\pm},y_{j_{2}}^{\pm})\no\\
&=G(Q_{1}-Q_{2})\kko{\prod_{n=1}^{Q_{2}-1}G(Q_{1}-Q_{2}+2n)^{2}}G(Q_{1}+Q_{2})\times {}\no\\[2mm]
& {}\qquad \times
\Sigma(Y_{1}^{\pm},Y_{2}^{\pm})^{-2}\,,\quad 
\mbox{where}\quad G(q)=\f{U_{1}-U_{2}-iq/2g}{U_{1}-U_{2}+iq/2g}\,.
\label{S-bounstates}
\end{align}
We note that $G(q)$ can be rewritten in terms of the spectral parameters as, 
\begin{equation}
G(q)
=\f{\ko{Y_{1}^{-}-Y_{2}^{+}}\ko{1-1/Y_{1}^{-}Y_{2}^{+}}+i(Q_{1}+Q_{2}-q)/2g}
	{\ko{Y_{1}^{+}-Y_{2}^{-}}\ko{1-1/Y_{1}^{+}Y_{2}^{-}}-i(Q_{1}+Q_{2}-q)/2g}\,.
\label{G(q)}
\end{equation}
Here $\Sigma^{-2}$ stands for the appropriate dressing factor for
boundstates. As explained in \cite{Chen:2006gq}, its form as a 
function of the spectral
parameters is identical to that of the conjectured BES dressing factor \cite{Beisert:2006ez}
for elementary magnons. Explicitly we have, 
\begin{align}
\Sigma(Y_{1}^{\pm},
Y_{2}^{\pm})^{-2}&=\prod_{j_{1}=1}^{Q_{1}}\prod_{j_{2}=1}^{Q_{2}}\sigma(y_{j_{1}}^{\pm},
y_{j_{2}}^{\pm})^{-2}=\ko{\f{R(Y_{1}^{+}, Y_{2}^{+})R(Y_{1}^{-},
    Y_{2}^{-})}
{R(Y_{1}^{+}, Y_{2}^{-})R(Y_{1}^{-}, Y_{2}^{+})}}^{-2}\,,
\label{dressbs}
\end{align}
where the final equality arises after numerous cancellations are taken
into account. 

\begin{figure}[htb]
\begin{center}
\vspace{.5cm}
\hspace{-.0cm}\includegraphics[scale=1.0]{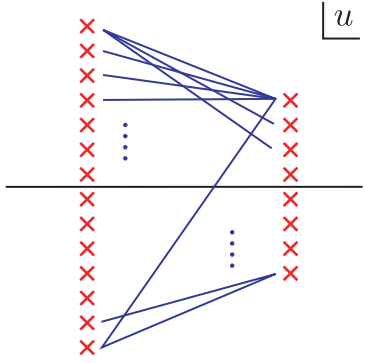}
\vspace{.0cm}
\caption{\small Constructing boundstate S-matrix by fusion.
Each boundstate is represented by an equally spaced sequence of Bethe roots (Bethe string).}
\label{fig:fusion}
\end{center}
\end{figure}
\paragraph{}
From Eqn (\ref{S-bounstates}) and (\ref{G(q)}), a finite set of simple
and double poles of the boundstate S-matrix are apparent. In addition,
as we review below, the dressing factor (\ref{dressbs}) provides an
infinite sequence of additional double poles. 
Below we will first investigate both simple and double poles of the
boundstate S-matrix (\ref{S-bounstates}) in turn, 
and discuss which of them are the physical singularities.
Then in Section \ref{sec:decode} we will interpret those
singularities as physical processes in terms 
of Landau diagrams.
We will also see their interpretation in terms Bethe root configurations in Section \ref{sec:Bethe String}.

\subsection{Simple Poles}

Simple poles are found in $G(Q_{1}+Q_{2})$ and $G(Q_{1}-Q_{2})$ at
\begin{equation}
U_{1}-U_{2}=-\f{i}{2g}(Q_{1}+Q_{2})
\quad \mbox{and}\quad 
U_{1}-U_{2}=-\f{i}{2g}(Q_{1}-Q_{2})\,.
\end{equation}
As we discuss below it is natural to interpret these poles as due to 
exchange of BPS boundstates of charge $Q_{1}\pm
Q_{2}$ in 
s- and t-channel
processes respectively\footnote{In the special case $Q_{1}=Q_{2}$\, 
the t-channel process is absent.}. 
Note however that, in terms of the spectral parameters $Y_{1}^{\pm}$
and $Y_{2}^{\pm}$ of the two incoming particles, 
$G(Q_{1}\pm Q_{2})$ are written as,
\begin{align}
G(Q_{1}+Q_{2})&=\f{Y_{1}^{-}-Y_{2}^{+}}{Y_{1}^{+}-Y_{2}^{-}}\cdot\f{1-1/Y_{1}^{-}Y_{2}^{+}}{1-1/Y_{1}^{+}Y_{2}^{-}}\,,\label{sp1}\\[2mm]
G(Q_{1}-Q_{2})&=\f{Y_{1}^{-}-Y_{2}^{-}}{Y_{1}^{+}-Y_{2}^{+}}\cdot\f{1-1/Y_{1}^{-}Y_{2}^{-}}{1-1/Y_{1}^{+}Y_{2}^{+}}\,,\label{sp2}
\end{align}
and, in these variables, there are two simple poles originating in
{\em each} of $G(Q_{1}+Q_{2})$ and
$G(Q_{1}-Q_{2})$.  
For example, $G(Q_{1}+Q_{2})$ has simple poles at
$Y_{1}^{+}=Y_{2}^{-}$ and 
$Y_{1}^{+}=1/Y_{2}^{-}$\,.
The question of which, if any, of these poles correspond to a physical
processes will be investigated in Section \ref{sec:physical poles}.

\subsection{Double Poles}
The singular structure of the dressing part is highly non-trivial, and
for scattering of elementary magnons, 
it was worked out in \cite{Dorey:2007xn}. This leads to an infinite
series of double poles in the magnon S-matrix. 
In the present
case of the 
boundstate S-matrix, there are two distinct sources of double poles 
which we will discuss in turn.

\subsubsection*{$\bullet$ BDS part}

The double poles of the BDS part of boundstate S-matrix
(\ref{S-bounstates}) 
locate at 
\begin{equation}
U_{1}-U_{2}=-\f{i}{2g}(Q_{1}+Q_{2}-2n)\,,\quad n=1,2,\dots, Q_{2}-1\,.
\label{u-u:3}
\end{equation}
As above, each of these, gives rise to a pair of double poles when
expressed in terms of the spectral parameters $Y_{1}^{\pm}$
and $Y_{2}^{\pm}$ at the two distinct roots of the equation\footnote{There are four ways to write down the condition (\ref{u-u:3}) in terms of the spectral parameters.
They all take the form $(Y_{1}^{\al}-Y_{2}^{\be})(1-1/Y_{1}^{\al}Y_{2}^{\be})+(i/g)\,n_{\al\be}=0$\,, where $(\al,\be)=(+,+), (+,-), (-,+), (-,-)$\,.
The integer $n_{\al\be}$ covers different region for these four choices, but the number of integers are the same, and is given by $\min\{Q_{1},Q_{2}\}-1$\,, namely $Q_{2}-1$ in our case.
The expression (\ref{u-u:3 x}) corresponds to $(\al,\be)=(+,-)$ case.
}
\begin{equation}
Y_{1}^{+}+\f{1}{Y_{1}^{+}}-Y_{2}^{-}-\f{1}{Y_{2}^{-}}-\f{in}{g}=0\,,\quad n=1,2,\dots, Q_{2}-1\,.
\label{u-u:3 x}
\end{equation}

\subsubsection*{$\bullet$ Dressing part}

As the functional form of the dressing factor is essentially the same as the
elementary magnon scattering case, we follow the analysis of 
\cite{Dorey:2007xn} .
In particular, we consider the derivative with respect to the coupling
$g$ of the function $\chi$ appearing in (\ref{dressbs}),  
\begin{align}
\pa_{g}\chi(Y_{1},Y_{2})&=-\oint_{\mathcal C}\f{dz_{1}}{2\pi}\oint_{\mathcal C}\f{dz_{2}}{2\pi}
\f{z_{1}+\f{1}{z_{1}}-z_{2}-\f{1}{z_{2}}}{(z_{1}-Y_{1})(z_{2}-Y_{2})}\,\komoji{\Psi\ko{1+ig\ko{z_{1}+\f{1}{z_{1}}-z_{2}-\f{1}{z_{2}}}}}\no\\[2mm]
&=-\oint_{\mathcal C}\f{dz_{1}}{2\pi}\oint_{\mathcal C}\f{dz_{2}}{2\pi}
\sum_{n=1}^{\infty}
\f{n/g^{2}}{(z_{1}-Y_{1})(z_{2}-Y_{2})}
\f{1}{z_{1}+\f{1}{z_{1}}-z_{2}-\f{1}{z_{2}}-\f{in}{g}}\,,
\label{del-g-chi}
\end{align}
where we used the definition of digamma function $\Psi(x)$ and its asymptotic expansion,
\begin{equation}
\f{d}{dx}\log \Gamma(x)=\Psi(x)
=-\gamma_{\rm E}-\sum_{n=1}^{\infty}\kko{\f{1}{x+n-1}-\f{1}{n}}
\end{equation}
($\gamma_{\rm E}$ is Euler's constant).
Let $z_{2}=F_{n}(z_{1})$ be the root of the quadratic equation
\begin{equation}
z_{1}+\f{1}{z_{1}}-z_{2}-\f{1}{z_{2}}-\f{in}{g}=0\,,
\end{equation}
which satisfies $|F_{n}(z_{1})|<1$\,. 
By the same argument as in \cite{Dorey:2007xn}, singularities arise
when poles of the integrand pinch the integration contour. 
As explained in Section 5 of\cite{Dorey:2007xn} the only case where we
pinch the contour is when 
$Y_{1}=Y_{1}^{-}$ and $Y_{2}=Y_{2}^{+}$\,.
Plugging this into (\ref{del-g-chi}) and performing the double contour integrals, we reach the expression
\begin{align}
\pa_{g}\chi(Y_{1}^{-},Y_{2}^{+})=
-\f{n}{g^{2}}\sum_{n=1}^{\infty}\f{F_{n}(Y_{1}^{-})}{\kko{Y_{2}^{+}F_{n}(Y_{1}^{-})-1}\kko{1-F_{n}(Y_{1}^{-})^{2}}}\,.
\end{align}
This can be easily integrated, giving
\begin{equation}
\chi(Y_{1}^{-},Y_{2}^{+})=-i\sum_{n=1}^{\infty}\log\ko{Y_{2}^{+}-F_{n}(Y_{1}^{-})^{-1}}\,.
\end{equation}
Then we see the relevant parts of our poles/zeros analysis become
\begin{equation}
\Sigma(Y_{1}^{\pm}, Y_{2}^{\pm})^{-2}
\sim e^{2i\kko{\chi(Y_{1}^{-}, Y_{2}^{+})-\chi(Y_{2}^{-}, Y_{1}^{+})}}
=\prod_{n=1}^{\infty}\kko{\f{Y_{2}^{+}-F_{n}(Y_{1}^{-})^{-1}}{Y_{1}^{+}-F_{n}(Y_{2}^{-})^{-1}}}^{2}\,.
\label{zero,pole}
\end{equation}
From (\ref{zero,pole}), we see the double poles lie at $Y_{1}^{+}=F_{n}(Y_{2}^{-})^{-1}$\,.
In view of $F_{n}(x)+F_{n}(x)^{-1}=x+x^{-1}-(in/g)$\,, this condition turns out 
\begin{equation}
Y_{1}^{+}+\f{1}{Y_{1}^{+}}-Y_{2}^{-}-\f{1}{Y_{2}^{-}}=-\f{in}{g}\,,
\label{u-u:2 x}
\end{equation}
which is one of the roots of the equation,  
\begin{equation}
U_{1}-U_{2}=-\f{i}{2g}(Q_{1}+Q_{2}+2n)\,,\quad n=1,2,\dots,\qquad U_{j}\eq u(Y_{j}^{\pm})\,.
\label{u-u:2}
\end{equation}
In the special case $Q_{1}=Q_{2}=1$, we reproduce the results of 
\cite{Dorey:2007xn}.

\section{Physicality Conditions\label{sec:physical poles}}

In general, S-matrix singularities occur at complex values of the external momenta and energies.
Only those singularities suitably close to the real axis (with positive energy) require a physical explanation.
In relativistic scattering there is a well-established notion of a ``physical sheet''.
In the present case, where the dynamics is non-relativistic, the
extent of the physical region is unclear. However, for each scattered
particle there are three distinct limits in which it is possible to
analyse the situation precisely. These are\,:
\begin{description}
\item[\bmt{(i)}] {\bf The Giant Magnon limit}\,:~ $g\to \infty$ while $P$ kept fixed, where 
\begin{equation}
Y^{+}\simeq 1/Y^{-}\simeq e^{iP/2}\,,\quad 
U\simeq 2\cos\ko{\f{P}{2}}\,,\quad 
E\simeq 4g\sin\ko{\f{P}{2}}\,.
\end{equation}
In this limit the particles with arbitrary charge $Q$ become heavy solitonic states of the
string worldsheet theory. 

\item[\bmt{(ii)}] {\bf Plane-Wave limit}\,:~ $g\to \infty$ with $k\eq 2gP$ kept fixed, where 
\begin{equation}
Y^{+}\simeq Y^{-}\simeq \f{Q+\sqrt{Q^{2}+k^{2}}}{k}\in {\mathbb R}\,,\quad 
U\simeq \f{2}{k}\sqrt{Q^{2}+k^{2}}\,,\quad 
E\simeq \sqrt{Q^{2}+k^{2}}\,.
\label{pp-wave region}
\end{equation}
In this limit the magnon reduces to an elementary excitation of the
worldsheet theory. As before states with $Q>1$ are interpreted as
boundstates of the elementary $Q=1$ excitation. 
Notice one can also express 
\begin{equation}
E=\f{\xi^{2}+1}{\xi^{2}-1}\,Q\,,\quad 
k=\f{2\xi}{\xi^{2}-1}\,Q\,,\quad 
\xi\,e^{\pm i\delta/2} \eq Y^{\pm}\quad (\xi\in\mathbb R\,,~ 0<\delta\ll 1)\,.
\label{pp-wave region 2}
\end{equation}

\item[\bmt{(iii)}] {\bf Heisenberg spin-chain limit}\,:~ $g\ll 1$ limit, where 
\begin{equation}
Y^{\pm}\mp \f{iQ}{2g}\simeq U\simeq \f{1}{2g}\cot\ko{\f{P}{2}}\,,\quad 
E\simeq Q+\f{8g^{2}}{Q}\sin^{2}\ko{\f{P}{2}}\,.
\end{equation}
In this limit, the gauge theory can be studied in the one-loop
approximation where the dilatation operator in the $SU(2)$ sector is
precisely the Heisenberg Hamiltonian. 

\end{description}

In the following, as in \cite{Dorey:2007xn}, we will focus on
singularities which lie 
parametrically close to the positive real axis 
for both external energies.
In particular, this includes those singularities which come close to
the positive real axis in any of the three limits described above. 
Below we will identify which poles of the boundstate S-matrix fall
into this category. We will refer to them as physical poles. 

\subsection{Physical Simple Poles}

In (\ref{sp1}), (\ref{sp2}), we saw there are two simple poles for each
of $G(Q_{1}+Q_{2})$ and $G(Q_{1}-Q_{2})$ 
when written in terms of the spectral parameters. We will study the
behaviour of these poles in the limits described above. The key
question is whether, in any of these limits, the pole approaches 
a point where the energies of both particles scattering are real and
positive. If this is the case in at least one of the limits considered
then we will accept the pole as physical.

\paragraph{$\bullet$ Simple poles in \bmt{G(Q_{1}+Q_{2})}.}

First consider the case where the momenta of the two external particles are in the plane-wave region.
This means we have $Y_{i}^{+}\simeq Y_{i}^{-}$ $(i=1,2)$\,.
Let us suppose the first particle ($i=1$) is in the physical region
(so its energy $E_{1}$ is positive), 
and see whether one of the pole conditions $Y_{1}^{+}=Y_{2}^{-}$ $(\eq e^{i\delta /2})$ implies the second particle ($i=2$) is also physical.
The answer can be found by looking at the relative sign of energies 
between the two particles.
Since $e^{-i\delta /2}\simeq Y_{1}^{-}\simeq Y_{1}^{+}=e^{i\delta
  /2}=Y_{2}^{-}\simeq Y_{2}^{+}\simeq e^{-i\delta /2}$\,, 
the energy of the second particle is evaluated as
\begin{align}
E_{2}&=\f{g}{i}\kko{\ko{Y_{2}^{+}-\f{1}{Y_{2}^{+}}}-\ko{Y_{2}^{-}-\f{1}{Y_{2}^{-}}}}\cr
&\simeq\f{g}{i}\kko{\ko{Y_{1}^{-}-\f{1}{Y_{1}^{-}}}-\ko{Y_{1}^{+}-\f{1}{Y_{1}^+}}}
= -E_{1} ~ <0\,,
\end{align}
thus the second particle does not live near the physical region in
this limit.
On the other hand, for the other pole at 
$Y_{1}^{+}=1/Y_{2}^{-}$\, we have $Y_{1}^{-}\simeq 1/Y_{2}^{+}$
in the plane-wave region which leads to $E_{2}>0$\, 
thus corresponding to physical pole.

Next let us consider the case of the scattering of two dyonic giant magnons.
In this case, the spectral parameters are related as 
$Y_{i}^{+}\simeq 1/Y_{i}^{-}$ $(i=1,2)$\,.
By insisting that the energy of both particles is positive in this
limit, we again select the pole at $Y_{1}^{+}=1/Y_{2}^{-}$. The other
pole at $Y_{1}^{+}=Y_{2}^{-}$ again violates this criterion.

Finally let us consider the Heisenberg spin-chain limit.
By noticing the $g$\,-dependence of the spectral parameters, RHS of (\ref{sp1}) becomes in this limit,
\begin{equation}
G(Q_{1}+Q_{2})=\f{Y_{1}^{-}-Y_{2}^{+}}{Y_{1}^{+}-Y_{2}^{-}}\cdot\ko{1+\ord{g^{2}}}\,,
\label{Heisenberg limit}
\end{equation}
so one finds, contrast to the strong coupling results, it is the pole
$Y_{1}^{+}=Y_{2}^{-}$ that 
should be regarded as physical pole in the weak coupling region. 
In fact we can write down the wavefunction of the corresponding
boundstate explicitly in this limit. 

In conclusion we have found at least one limit in which each of the
two simple poles (at $Y_{1}^{+}=Y_{2}^{-}$ and at
$Y_{1}^{+}=1/Y_{2}^{-}$) occurs near the region of positive real
energies. Thus we will accept both poles as physical and seek an
explanation in terms of on-shell intermediate states. 

\paragraph{$\bullet$ Simple poles in \bmt{G(Q_{1}-Q_{2})}.}

We can apply the same line of reasoning to this case.
In particular one can show that the pole at $Y_{1}^{+}=Y_{2}^{+}$\,, 
occurs near the region where both external particles have real
positive energies, in all the three of the limits discussed above 
(giant magnon, plane wave and Heisenberg spin-chain). In contrast one may
check that the remaining pole at $Y_{1}^{+}=1/Y_{2}^{+}$ stays away
from the physical region in each of the limits considered. For this
reason we will accept the first pole as physical but not the second. 

\paragraph{}
In summary, three of the four poles, $Y_{1}^{+}=Y_{2}^{-}$\,, $Y_{1}^{+}=1/Y_{2}^{-}$ and $Y_{1}^{+}=Y_{2}^{+}$ are identified with physical poles, giving $E_{i}>0$ and $Q_{i}>0$ for both external particles $Y_{i}$ ($i=1,2$) at least one of the three $(i)$-$(iii)$ regions.
They are summarised in Table \ref{tab:physical simple poles}.
Entries with checks ``\check\,'' indicate the pole result in $E_{i}>0$ and $Q_{i}>0$ in the region for both $i=1,2$\,.

\begin{table}[htbp]
\caption{\small The first three poles ($Y_{1}^{+}=Y_{2}^{-}$\,, $Y_{1}^{+}=1/Y_{2}^{-}$\,, $Y_{1}^{+}=Y_{2}^{+}$) are physical while the other one ($Y_{1}^{+}=1/Y_{2}^{+}$) is unphysical.}
\begin{center}
\begin{tabular}{|cl||c|c||c|c|}\hline
{} &  & \multicolumn{2}{c||}{$G(Q_{1}+Q_{2})$} & \multicolumn{2}{c|}{$G(Q_{1}-Q_{2})$}\\ \cline{3-6}
{} & {} & $Y_{1}^{+}=Y_{2}^{-}$ & $Y_{1}^{+}=1/Y_{2}^{-}$ & $Y_{1}^{+}=Y_{2}^{+}$ & $Y_{1}^{+}=1/Y_{2}^{+}$ \\ \hline
$(i)$ & Giant Magnon limit & \batsu & \check & \check & \batsu \\
$(ii)$ & Plane-Wave limit & \batsu & \check & \check & \batsu \\
$(iii)$ & Heisenberg spin-chain limit & \check & \batsu & \check & \batsu \\ \hline
\end{tabular}
\end{center}
\label{tab:physical simple poles}
\end{table}

\subsection{Physical Double Poles}

As we saw in the previous sections, double poles exist in two regions\,; one is in a finite interval (\ref{u-u:3 x}) that comes from the BDS part, and the other is an infinite interval (\ref{u-u:2 x}) from the dressing part.

\paragraph{The BDS part.}
We start with investigating the first region originated from the BDS part.
Written in terms of the spectral parameters, there are two double poles in the BDS part of boundstate S-matrix (\ref{S-bounstates}), which are the two roots of the equation (\ref{u-u:3 x}).
For each $n$\,, one can solve the constraint for $Y_{1}^{+}$ to find the two roots
\begin{equation}
Y_{1}^{+}=y_{\pm}^{(n)}\eq \f{g(Y_{2}^{-})^{2}+i n Y_{2}^{-}+g\pm\sqrt{\ko{g(Y_{2}^{-})^{2}+i n Y_{2}^{-}+g}^{2}-4g^{2}(Y_{2}^{-})^{2}}}{2gY_{2}^{-}}\,.
\label{y pm}
\end{equation}
Here the subscripts in $y_{\pm}^{(n)}$ refers to the signs in front of the square root in (\ref{y pm}), and the integer $n$ runs $n=1,\dots,Q_{2}-1$\,.
We would like to find out which of the two roots corresponds to a physical double pole that satisfy the criteria established in the previous section.
We will do the examination for $n\sim 1$ and $n\sim Q_{2}$ regions, separately, when the identification becomes transparent due to that we already know which are the closest physical simple poles to each regions.
For the latter case with $n\sim Q_{2}$\,, we will further divide the case into two according to whether $Q_{2}\ll g$ or $Q_{2}\gg g$\,.

\paragraph{}
When $n$ is much smaller than $g$\,, the two roots in (\ref{y pm}) approach to $y_{+}\to Y_{2}^{-}$ and $y_{-}\to 1/Y_{2}^{-}$\,.
In this region, since $n$ is close to zero, we can expect the physical double poles exist near the physical simple pole in $G(Q_{1}+Q_{2})$\,, which is $Y_{1}^{+}=1/Y_{2}^{-}$ as we identified in the previous section (see Table \ref{tab:physical simple poles}).
Hence we conclude that in both the plane wave and the giant magnon regions, it is the root $y_{-}$ that corresponds to a physical double pole, while in the Heisenberg spin-chain limit $g\ll 1$\,, the other root $y_{+}$ is physical since $y_{-}$ disappears in this limit, just as we saw in (\ref{Heisenberg limit}).

Next let us turn to the other side of the BDS double pole spectrum, $n=Q_{2}-1,\,Q_{2}-2,\dots$\,, which is near the physical simple pole in $G(Q_{1}-Q_{2})$\,.
First notice that when $n$ is close to $Q_{2}$\,, the two roots in (\ref{y pm}) approach to either $Y_{2}^{+}$ or $1/Y_{2}^{+}$\,.
To see this, let us solve the constraint $Y_{2}^{+}+1/Y_{2}^{+}-iQ_{2}/2g=Y_{2}^{-}+1/Y_{2}^{-}+iQ_{2}/2g$ for $Y_{2}^{+}$ \,, which leads to
\begin{equation}
Y_{2}^{+}=y'{}_{\pm}\eq \f{g(Y_{2}^{-})^{2}+i Q_{2} Y_{2}^{-}+g\pm\sqrt{\ko{g(Y_{2}^{-})^{2}+i Q_{2} Y_{2}^{-}+g}^{2}-4g^{2}(Y_{2}^{-})^{2}}}{2gY_{2}^{-}}\,.
\label{y' pm}
\end{equation}
Comparing them with $y_{\pm}^{(n)}$ in (\ref{y pm}), we see that as $n$ tends to $Q_{2}$\,, two branches $y_{\pm}^{(n)}$ approach $y'{}_{\pm}$ respectively.
Recall that in the previous section we found the only physical simple pole of $G(Q_{1}-Q_{2})$ in all three regions (the plane wave, giant magnon and Heisenberg regions) was $Y_{1}^{+}=Y_{2}^{+}$\,, and the physical double poles with $n=Q_{2}-1,\,Q_{2}-2,\dots$ should be close to it.
These observations lead us to conclude that, for $n$ close to $Q_{2}$\,, if $Y_{2}^{+}=y'{}_{+}$\,, the physical double pole is given by $Y_{1}^{+}=y_{+}^{(n)}$\,, whereas if $Y_{2}^{+}=y'{}_{-}$\,, it is given by $Y_{1}^{+}=y_{-}^{(n)}$\,, for all the three regions.

Which of $y_{\pm}^{(n)}$ should be singled out as the physical branch of $Y_{1}^{+}$ around $n\sim Q_{2}$ depends on the magnitude of $Q_{2}$ compared to $g$\,.
When $Q_{2}\ll g$\,, the two branches $y'{}_{+}$ and $y'{}_{-}$ approach $Y_{2}^{-}$ and $1/Y_{2}^{-}$\,, respectively.
Therefore, for the plane wave region where $Y_{2}^{+}\simeq Y_{2}^{-}$\,, we should single out $Y_{2}^{+}=y'{}_{+}$ so that the physical double pole corresponds to $Y_{1}^{+}=y_{+}^{(n)}$\,.
On the other hand, in the giant magnon region where $Y_{2}^{+}\simeq 1/Y_{2}^{-}$\,, we should single out $Y_{2}^{+}=y'{}_{-}$ so that the physical double pole corresponds to $Y_{1}^{+}=y_{-}^{(n)}$\,.
As for the Heisenberg spin-chain limit, there is no such limit we can take for the current $Q_{2}\ll g$ case.
The results for $Q_{2}\ll g$ case is summarised in Table \ref{tab:physical double poles 1}.
In the giant magnon region, the physical double pole remains to be $Y_{1}^{+}=y_{-}^{(n)}$ for all $n=1,\dots,Q_{2}-1$\,, while in the the plane wave region, the physical double pole switches from $Y_{1}^{+}=y_{-}^{(n)}$ ($n\sim 1$) to $Y_{1}^{+}=y_{+}^{(n)}$ ($n\sim Q_{2}$) around some point.

\begin{table}[htbp]
\caption{\small The $Q_{2}\ll g$ case.}
\begin{center}
\begin{tabular}{|cl||c|c||c|c|}\hline
{} &  & \multicolumn{2}{c||}{$n\sim 1$} & \multicolumn{2}{c|}{$n\sim Q_{2}\, (\ll g)$}\\ \cline{3-6}
{} & {} & $Y_{1}^{+}=y_{+}^{(n)}$ & $Y_{1}^{+}=y_{-}^{(n)}$ & $Y_{1}^{+}=y_{+}^{(n)}$ & $Y_{1}^{+}=y_{-}^{(n)}$ \\ \hline
$(i)$ & Giant Magnon limit & \batsu & \check & \batsu & \check \\
$(ii)$ & Plane-Wave limit & \batsu & \check & \check & \batsu \\
$(iii)$ & Heisenberg spin-chain limit & - & - & - & - \\ \hline
\end{tabular}
\end{center}
\label{tab:physical double poles 1}
\end{table}

\paragraph{}
When $Q_{2}\gg g$\,, the two branches $y'{}_{+}$ and $y'{}_{-}$ reduce to $iQ_{2}/g$ and $0$\,, respectively.
Hence the physical double poles around $n\sim Q_{2}$ are singled out to be $Y_{1}^{+}=y_{+}^{(n)}$ corresponding to the $y'{}_{+}$ branch of $Y_{2}^{+}$\,, since the other root $y_{-}^{(n)}$ disappears just like the case with the Heisenberg spin-chain limit (\ref{Heisenberg limit}).
As a result, in all the three regions, we conclude that $Y_{1}^{+}=y_{+}^{(n)}$ corresponds to the physical double pole when $n$ is close to $Q_{2}\, (\gg g)$\,.
The results are summarised in Table \ref{tab:physical double poles 2}.

\begin{table}[htbp]
\caption{\small The $Q_{2}\gg g$ case.}
\begin{center}
\begin{tabular}{|cl||c|c||c|c|}\hline
{} &  & \multicolumn{2}{c||}{$n\sim 1$} & \multicolumn{2}{c|}{$n\sim Q_{2}\, (\gg g)$}\\ \cline{3-6}
{} & {} & $Y_{1}^{+}=y_{+}^{(n)}$ & $Y_{1}^{+}=y_{-}^{(n)}$ & $Y_{1}^{+}=y_{+}^{(n)}$ & $Y_{1}^{+}=y_{-}^{(n)}$ \\ \hline
$(i)$ & Giant Magnon limit & \batsu & \check & \check & \batsu \\
$(ii)$ & Plane-Wave limit & \batsu & \check & \check & \batsu \\
$(iii)$ & Heisenberg spin-chain limit & \check & \batsu & \check & \batsu \\ \hline
\end{tabular}
\end{center}
\label{tab:physical double poles 2}
\end{table}

\paragraph{The dressing part.}
Next let us turn to the double poles from the dressing phase, (\ref{u-u:2 x}).
Actually the analysis for this case is already basically done, since (\ref{u-u:2 x}) leads to the same equation (\ref{y pm}).
The only difference lies in the range of $n$\,, which runs $n=Q_{2}+1,\,Q_{2}+2,\dots$ in this case.
Therefore, in order to identify the physical double poles, we have only to refer to Tables \ref{tab:physical double poles 1} and \ref{tab:physical double poles 2}.
In the giant magnon region, $Y_{1}^{+}=y_{-}^{(n)}$ corresponds to the physical double pole when $n\ll g$\,, while when $n\gg g$\,, the other branch $Y_{1}^{+}=y_{+}^{(n)}$ plays the role.
In the plane wave region, $Y_{1}^{+}=y_{+}^{(n)}$ remains the physical double pole for all $n=Q_{2}+1,\,Q_{2}+2,\dots$\,.

\section{Decoding Physical Poles\label{sec:decode}}

For the conjectured boundstate S-matrix to be correct, there should
exist at least one physical process that 
accounts for each physical pole.
In other words, we should be able to draw at least one consistent
Landau diagram. 
Also, there should not be any Landau diagrams which lead to extra
poles in the physical region which are not seen in the S-matrix. 
In this section, we will draw Landau diagrams corresponding to the
physical poles we identified above and comment on the possible
occurrence of other diagrams.

\begin{figure}[tb]
\begin{center}
\vspace{.5cm}
\hspace{-.0cm}\includegraphics[scale=0.7]{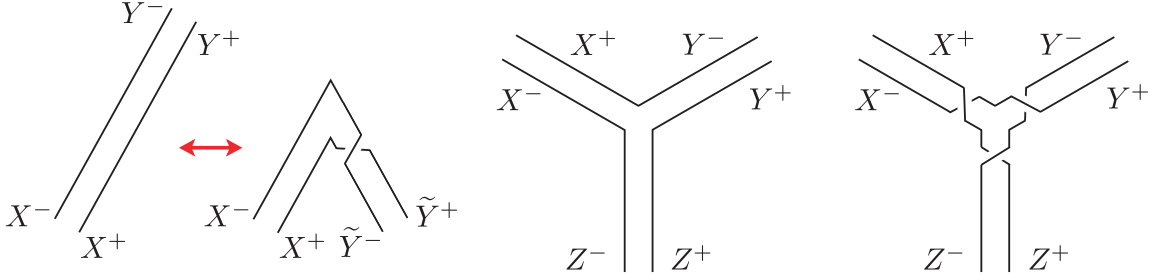}
\vspace{.0cm}
\caption{\small Building blocks of physical processes.
The ``double line'' notation of \cite{Dorey:2007xn} is employed, and time flows from bottom to top.
The dotted line indicates the corresponding particles carry negative charges.
}
\label{fig:Landau}
\end{center}
\end{figure}

The rules for constructing these diagrams are the same as given
in \cite{Dorey:2007xn}. Our current analysis generalises that of \cite{Dorey:2007xn} in that
we are analysing the situation where 
both the external (incoming/outgoing) particles carry generic (positive) charges, which we denote as $Q_{1}$ and $Q_{2}$ ($Q_{1}\geq Q_{2}$).
The building blocks of physical processes are the three particle
vertices shown in Figure
\ref{fig:Landau} which implement conservation of energy, momentum and
other quantum numbers. The left diagram shows the crossing transformation, $\widetilde Y^{\pm}=1/Y^{\pm}$\,. 
The other two diagrams describe two possible three-vertex diagrams.
The spectral parameters of the three particles
are related as 
$X^{+}=Y^{-}$\,, $X^{-}=Z^{-}$ and $Y^{+}=Z^{+}$ for the middle, and 
$X^{+}=Z^{+}$\,, $X^{-}=Y^{+}$ and $Y^{-}=Z^{-}$ for the right. All
lines in the diagram are on-shell. 

\subsection{Landau diagrams for simple poles}

As we saw in the previous section, there are three physical simple poles, and there must be at least one corresponding Landau diagram for each of them.
Let us first forget about the physicality condition and try to draw down the diagrams endowed with four simple pole conditions $Y_{1}^{+}=Y_{2}^{-}$ and $Y_{1}^{+}=1/Y_{2}^{-}$ (from $G(Q_{1}+Q_{2})$) and $Y_{1}^{+}=Y_{2}^{+}$ and $Y_{1}^{+}=1/Y_{2}^{+}$ (from $G(Q_{1}-Q_{2})$).
In any case, those simple poles describe formations of boundstate $Z^{\pm}$ either in the s- or t-channel.
Let us denote the multiplet number of the intermediate BPS particle $Q_{Z}$\,, which can be found out by the formula
\begin{equation}
Q_{Z}=\f{g}{i}\kko{\ko{Z^{+}+\f{1}{Z^{+}}}-\ko{Z^{-}+\f{1}{Z^{-}}}}\,.
\label{Qz}
\end{equation}
Since we assumed $Q_{1}>Q_{2}$\,, the multiplet number $Q_{Z}\, (>0)$ can only take values either $Q_{Z}=Q_{1}+Q_{2}$ or $Q_{Z}=Q_{1}-Q_{2}$\,, and the $U(1)$ charge carried by $Z^{\pm}$ is $Q_{1}+Q_{2}$ if the process is in the s-channel, and $Q_{1}-Q_{2}$ or $-(Q_{1}-Q_{2})$ if it is in the t-channel.
Notice in our convention the multiplet number $Q_{Z}$ must be positive while the $U(1)$ charge can take either positive or negative values, varying from $-Q_{Z}$ to $+Q_{Z}$\,.

\begin{figure}[tb]
\begin{center}
\vspace{.5cm}
\hspace{-.0cm}\includegraphics[scale=0.7]{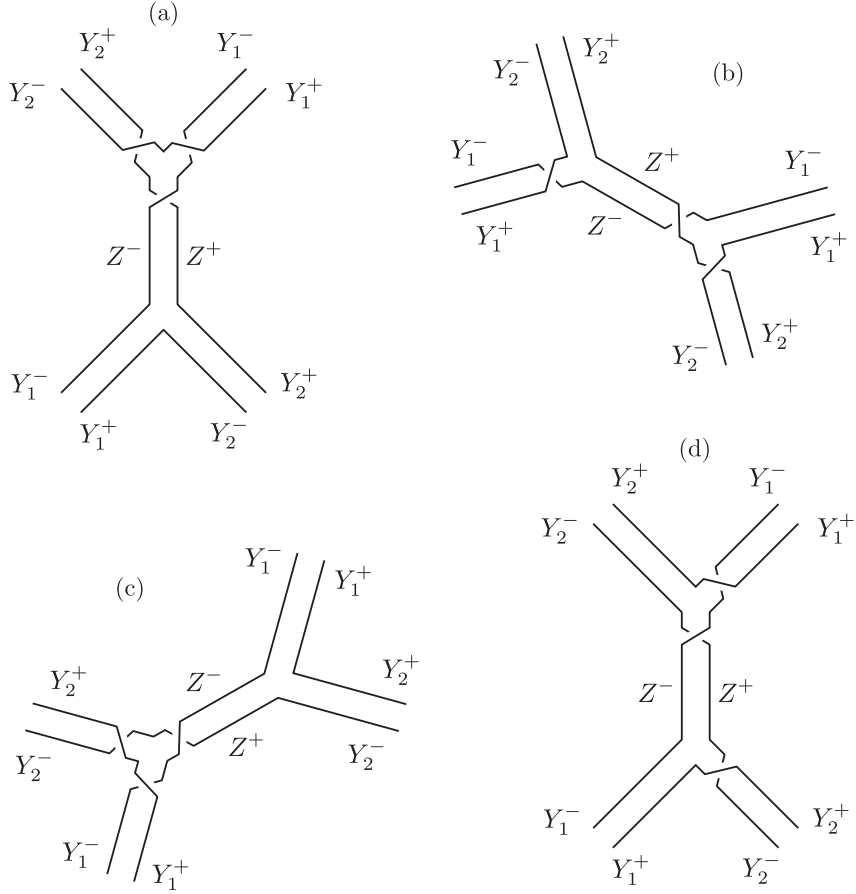}
\vspace{.0cm}
\caption{\small (Examples of) diagrams describing four simple poles $Y_{1}^{+}=Y_{2}^{-}$\,, $Y_{1}^{+}=1/Y_{2}^{-}$\,, $Y_{1}^{+}=Y_{2}^{+}$ and $Y_{1}^{+}=1/Y_{2}^{+}$\,.
They correspond to the diagrams (a)\,-\,(d) respectively.
The diagrams (a)\,-\,(c) describe physical processes, whereas (d) is not an allowed process.
}
\label{fig:simple}
\end{center}
\end{figure}

One can draw Landau diagrams corresponding to the poles $Y_{1}^{+}=Y_{2}^{-}$ and $Y_{1}^{+}=1/Y_{2}^{+}$ uniquely, which are shown in Figure \ref{fig:simple} (a) and (d), respectively.
In both cases the intermediate particle belongs to the multiplet $Q_{Z}=Q_{1}+Q_{2}$\,.
As for the rest two poles, for each $Y_{1}^{+}=1/Y_{2}^{-}$ and $Y_{1}^{+}=Y_{2}^{+}$\,, there are two diagrams possible\,;
One of them corresponds to the case where $U(1)$ charge carried by $Z^{\pm}$ is positive, while the other it is negative.
Still, in both cases the intermediate particle belongs to the multiplet $Q_{Z}=Q_{1}-Q_{2}>0$\,.
For each of these two simple poles, only one of the two possibilities is displayed in Figure \ref{fig:simple} (b) and (c), such that in (b) the $U(1)$ charge of $Z^{\pm}$ is negative, while in (c) it is positive.

The pole conditions, the multiplet number and the $U(1)$ charge of the intermediate particle $Z^{\pm}$ associated with the Landau diagrams (a)\,-\,(d) in Figure \ref{fig:simple} are summarised in Table \ref{tab:a-d}.
For example, for the boundstate formation process (a), by plugging the pole condition $Y_{1}^{+}=Y_{2}^{-}$ and the other constraints $Y_{1}^{-}=Z^{-}$\,, $Y_{2}^{+}=Z^{+}$ into (\ref{Qz}), one finds $Q_{Z}=Q_{1}+Q_{2}$\,, and the $U(1)$ charge carried by $Z^{\pm}$ is $Q_{1}+Q_{2}$ since it is in the s-channel.
The rest diagrams can be worked out in the same way.

\begin{table}[htbp]
\caption{\small Four simple poles and corresponding diagrams.
The first three (a)\,-\,(c) are physical process while the last (d) is not allowed.}
\begin{center}
\begin{tabular}{|c|c|c|c|c|c|}\hline
& simple pole & constraints & $Q_{Z}$ & charge of $Z^{\pm}$ & physcality \\ \hline
(a) & $Y_{1}^{+}=Y_{2}^{-}$ \hfill & $Y_{1}^{-}=Z^{-}\,,~ Y_{2}^{+}=Z^{+}$ \hfill & $Q_{1}+Q_{2}$ & $Q_{1}+Q_{2}$ & \check\\
(b) & $Y_{1}^{+}=1/Y_{2}^{-}$ \hfill & $Y_{1}^{-}=1/Z^{-}\,,~ Y_{2}^{+}=Z^{+}$ \hfill & $Q_{1}+Q_{2}$ & $Q_{1}-Q_{2}$ & \check\\
(c) & $Y_{1}^{+}=Y_{2}^{+}$ \hfill & $Y_{1}^{-}=Z^{-}\,,~ Y_{2}^{-}=Z^{+}$ \hfill & $Q_{1}-Q_{2}$ & $Q_{1}-Q_{2}$ & \check\\
(d) & $Y_{1}^{+}=1/Y_{2}^{+}$ \hfill & $Y_{1}^{-}=Z^{-}\,,~ Y_{2}^{-}=1/Z^{+}$ \hfill & $Q_{1}-Q_{2}$ & $Q_{1}+Q_{2}$ & \batsu \\ \hline
\end{tabular}
\end{center}
\label{tab:a-d}
\end{table}

We can now see that the case (d) is impossible since it corresponds to a process where the intermediate particle belongs to multiplet $Q_{1}-Q_{2}$ but has $U(1)$ charge $Q_{1}+Q_{2}$\,.
This fact indicates the simple pole $Y_{1}^{+}=Y_{2}^{+}$ is not a physical pole, which is consistent with what we found in the previous section.

\subsection{Landau diagrams for double poles}

The relevant diagrams are the ``box'' and ``bow-tie'' diagrams, 
which were studied in \cite{Dorey:2007xn} for the elementary magnon scattering case.

\subsubsection{``Box'' diagram}

There are two possibilities here concerning the charges of the intermediate states;
\begin{description}
\item[\rm Case (A)\,:]
Both intermediate particles carry positive charges.
\item[\rm Case (B)\,:]
One of them carries positive charge while the other negative.
\end{description}
We will examine both cases in turn, and see they give rise to double
poles in two 
complementary regions in the parameter space.

\paragraph{$\bullet$ Double poles in Case (A).}

\begin{figure}[tb]
\begin{center}
\vspace{.5cm}
\hspace{-.0cm}\includegraphics[scale=0.7]{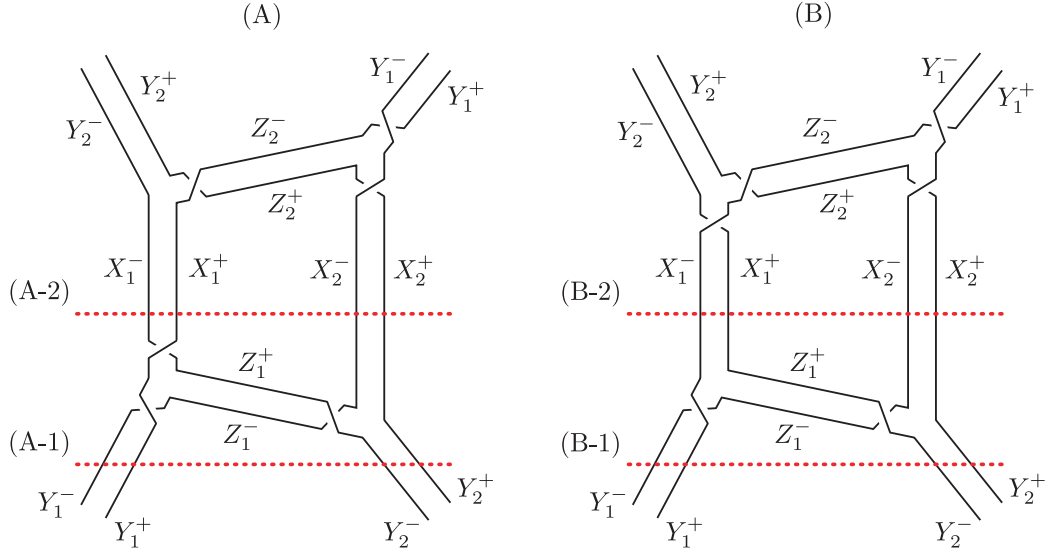}
\vspace{.0cm}
\caption{\small ``Box'' processes that give rise to double poles.
In (A), the absolute value of $X_{1}^{\pm}$ is greater than one, while in (B) it is smaller than one (see Figure \ref{fig:rap}).
In Section \ref{sec:Bethe String}, the two sets of time-slices, (A-1,2) and (B-1,2), will be interpreted as two different ways of viewing the same Bethe root configurations.
}
\label{fig:dp}
\end{center}
\end{figure}

The corresponding box diagram is shown in Figure \ref{fig:dp} (A).
We assigned spectral parameters $Y_{1}^{\pm}$ and $Y_{2}^{\pm}$ to the
two external particles, and 
$X_{1}^{\pm}$ and $X_{2}^{\pm}$ to the intermediate particles.
When the particle with $X_{1}^{\pm}$ carries positive charge $m$\,,
the two exchanged particles with spectral parameters $Z_{1}^{\pm}$ and
$Z_{2}^{\pm}$\,, carry negative charges $-(Q_{1}-m)$ and
$-(Q_{2}-m)$\,, respectively, 
in view of the charge conservation.
Here $m$ takes values $m=1,2,\dots, Q_{2}-1$ (we assumed $Q_{1}\geq Q_{2}$ as before).
Further by taking into account for the conservation of energy and
momentum at all vertices, 
one can show the spectral parameters must satisfy 
\begin{alignat}{3}
X_{2}^{+}&=Y_{2}^{+}=1/Z_{2}^{+}\,,&\qquad X_{2}^{-}&=Y_{1}^{-}=1/Z_{1}^{-}\,,\label{XYZ1}\\
X_{1}^{-}&=Y_{2}^{-}=1/Z_{1}^{+}\,,&\qquad X_{1}^{+}&=Y_{1}^{+}=1/Z_{2}^{-}\label{XYZ2}\,.
\end{alignat}
Using (\ref{XYZ1}), it is easy to verify that in this case (A) the double poles locate at
\begin{align}
U_{1}-U_{2}
&=\f{1}{Z^{+}}+Z^{+}-\f{1}{Z^{-}}-Z^{-}\cr
&=-\f{i}{2g}(Q_{1}+Q_{2}-2m)\,,\quad m=1,2,\dots, Q_{2}-1\,.
\label{u-u:1'}
\end{align}
where as before $U_{j}\eq u(Y_{j}^{\pm})$\,.
We see the number of double poles $Q_{2}-1$ is finite in this Case
(A), and the location exactly matches 
with the double poles in the BDS part of conjectured boundstate S-matrix, given in (\ref{u-u:3}).

\paragraph{$\bullet$ Double poles in region (B).}

The Figure \ref{fig:dp} (B) shows the box diagram process of Case (B),
where the particle with spectral parameter $X_{1}^{\pm}$ 
carries negative charge $-m<0$\,.
We assigned all the spectral parameters as the same as Case (A).
Then the energy and momentum conservation at all vertices imply, for one condition, the same as (\ref{XYZ1}), while on the other, 
\begin{alignat}{3}
1/X_{1}^{+}&=Y_{2}^{-}=1/Z_{1}^{+}\,,&\qquad 1/X_{1}^{-}&=Y_{1}^{+}=1/Z_{2}^{-}\label{XYZ22}
\end{alignat}
instead of (\ref{XYZ2}).
We see the spectral parameters $X_{1}^{\pm}$ in (\ref{XYZ2}) has replaced with $1/X_{1}^{\mp}$ in (\ref{XYZ22}), which is just the combination of the maps (\ref{interchange}) and (\ref{crossing}) that only flips the sign of the charge, unchanging energy and momentum.
Since now the two exchanged particles with $Z_{1}^{\pm}$ and $Z_{2}^{\pm}$ carry negative charges $-(Q_{1}+m)$ and $-(Q_{2}+m)$ respectively, the locations of the double poles become, in light of (\ref{XYZ1}), 
\begin{equation}
U_{1}-U_{2}
=-\f{i}{2g}(Q_{1}+Q_{2}+2m)\,,\quad m=1,2,\dots\,.
\label{u-u:1}
\end{equation}
This is an infinite series, and matches with the location of the double poles in the BES dressing part of conjectured 
boundstate S-matrix, given in (\ref{u-u:2}). 
The situation considered in \cite{Dorey:2007xn} corresponds to $Q_{1}=Q_{2}=1$ case.
Note also there is no double pole at
$U_{1}-U_{2}=-i(Q_{1}+Q_{2})/(2g)$\,; instead there is a simple pole
there, corresponding a formation of boundstate with charge
$Q_{1}+Q_{2}$ in the s-channel process shown in Figure \ref{fig:simple} (a).

\paragraph{}
Let us summarise.
For the box diagram case, double poles are found in two separate regions for given $Q_{1}$ and $Q_{2}$ (with $Q_{1}\geq Q_{2}$), as
\begin{align}
&U_{1}-U_{2}=-\f{i}{2g}(Q_{1}+Q_{2}+2n)\,,\no\\
&\qquad \mbox{where}\quad 
n=
\left\{
\begin{array}{ll}
\ds -Q_{2}+1\,, \dots\,, -2\,, -1 & \mbox{for Case (A)}\,, \\
\ds 1\,, 2\,, \dots & \mbox{for Case (B)}\,.
\end{array}
\right.
\label{double poles}
\end{align}
The poles in (A) originate from the BDS part of the boundstate
S-matirix, whereas the ones in (B) comes from the dressing
factor. 
Each of these equations (\ref{double poles}) has two roots, and which of them corresponds to the physical double pole is summarised in Tables  \ref{tab:physical double poles 1} and \ref{tab:physical double poles 2}.

\subsubsection{``Bow-Tie'' diagram}

\begin{figure}[htb]
\begin{center}
\vspace{.5cm}
\hspace{-.0cm}\includegraphics[scale=0.6]{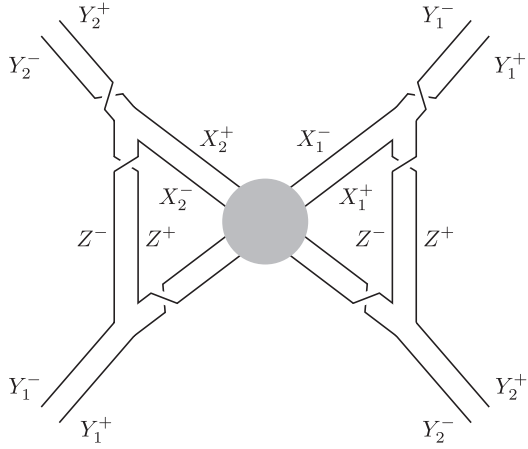}
\vspace{.0cm}
\caption{\small ``Bow-tie'' process that also rises to double poles \cite{Dorey:2007xn}.
There is a blob in the centre of the diagram, therefore special care has to be paid for the case both the intermediate plane wave magnons carry charge $Q_{1}+Q_{2}$\,.
}
\label{fig:wg}
\end{center}
\end{figure}

Generalisation of the other bow-tie shaped diagram to the boundstate scattering case is also straightforward.
Setting both the charges carried by the two intermediate giant magnons as $m$\,, the charges of plane wave magnons exchanged by the giants are $Q_{1}-m$ and $Q_{2}-m$\,, which are both negative. 
In the same manner as in the ``box'' diagram case, by using the relations
\begin{alignat}{3}
1/X_{1}^{+}&=Y_{2}^{+}=Z^{+}\,,&\qquad 1/X_{2}^{-}&=Y_{1}^{-}=Z^{-}\,,\\
1/X_{2}^{+}&=Y_{2}^{-}\,,&\qquad 1/X_{1}^{-}&=Y_{1}^{+}\,,
\end{alignat}
one can show that, setting $Q_{2}=\min\{Q_{1},Q_{2}\}$\,, the double poles apparently locate at
\begin{equation}
U_{1}-U_{2}= -\f{i}{2g}(Q_{1}+Q_{2}+2n)\,,\quad 
\mbox{with}\quad 
n\geq -Q_{2}+1\,.
\end{equation}
However, as in the case of \cite{Dorey:2007xn}, one has to take account of the effect of the blob at the centre of the diagram.
When $m=Q_{1}+Q_{2}$\,, the blob corresponds to the scattering of two anti-magnons with charges $-Q_{1}$ and $-Q_{2}$ with spectral parameters $X_{1}^{\pm}$ and $X_{2}^{\pm}$\,.
Then by considering the self-consistency condition of the diagram, the degree of the pole with $m=Q_{1}+Q_{2}$ turn out not two but one.
Therefore, again, there is a gap in the spectrum at $U_{1}-U_{2}=-i(Q_{1}+Q_{2})/(2g)$\,, and leads to the same spectrum as the ``box'' case (\ref{double poles}).

\subsubsection{Other diagrams}

We have so far been able to account for all the physical simple and double poles in the conjectured boundstate S-matrix, by finding (at least one) Landau diagrams for them.
Finally let us note that one can also 
draw lots of Landau diagrams leading to unphysical poles.
It is meaningful to note that they may or may not match the 
unphysical poles of the boundstate S-matrix.
It is also possible those diagrams which do not satisfy the physicality conditions lead to a set of double poles that coincides with the physical double poles (\ref{double poles}).
For example, one can draw ``sandglass'' shaped Landau diagrams which are obtained by rotating the bow-tie diagram by $90^{\circ}$\,.
There are many sandglass diagrams where all charges are conserved at
each vertex, but which do not satisfy the extra physicality
conditions. Some of these diagrams give rise to the same set of 
double poles as the physical ones.

\section{Bethe String Interpretation\label{sec:Bethe String}}

In this section we are going to discuss how two time-slices (A-1,2) and (B-1,2) in Figure \ref{fig:dp} (the ``box'' diagram), which correspond to external and the internal on-shell states respectively, are interpreted as two different ways of viewing the same Bethe root configurations. 
In terms of the root configurations, the origins of double poles in both the BDS and the dressing pieces can be understood intuitively.

\subsubsection*{BDS part\,:~double poles from ``overlaps''}

\begin{figure}[tb]
\begin{center}
\vspace{.5cm}
\hspace{-.0cm}\includegraphics[scale=0.95]{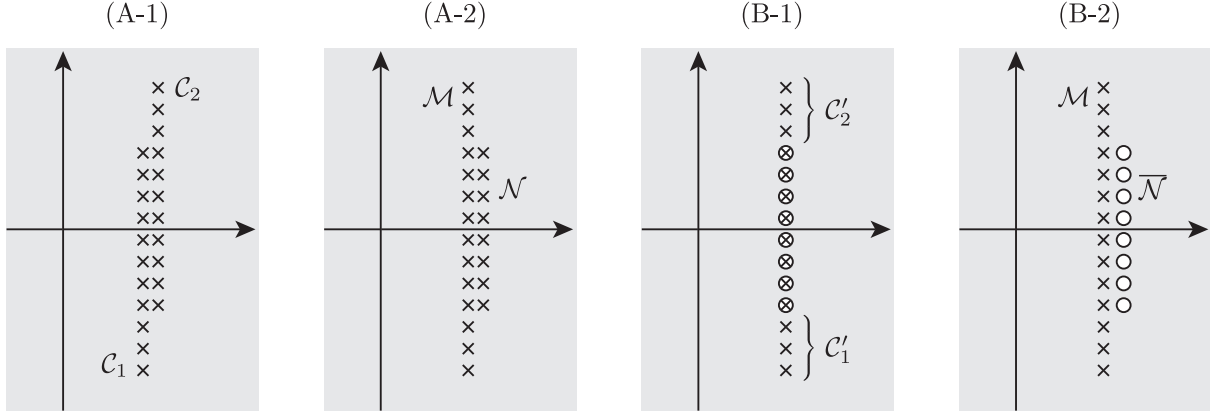}
\vspace{.0cm}
\caption{\small {[A-1]:} Bethe root configuration describing the two external particles in Figure \ref{fig:dp} (A).
\small {[A-2]:} The two intermediate BPS particles in Figure \ref{fig:dp} (A), both $\mathcal M$ and $\mathcal N$ carrying positive charges.
{[B-1]:} The two external particles in Figure \ref{fig:dp} (B).
\small {[B-2]:} The two intermediate BPS particles in Figure \ref{fig:dp} (B), where $\mathcal M$ carries positive charge while $\overline{\mathcal N}$ carries negative charge.}
\label{fig:rap}
\end{center}
\end{figure}

Let us first see the origin of the double poles (\ref{u-u:3}) in the rapidity plane.
Actually the same Bethe root configuration describing the double poles (\ref{u-u:3}) can be interpreted in two ways, each corresponding to two time-slices (A-1,2) of Figure \ref{fig:dp} (A).
The root configurations corresponding to these time-slices are shown in Figure \ref{fig:rap} (A-1,2), respectively.

\begin{figure}[tb]
\begin{center}
\vspace{.5cm}
\hspace{-.0cm}\includegraphics[scale=0.95]{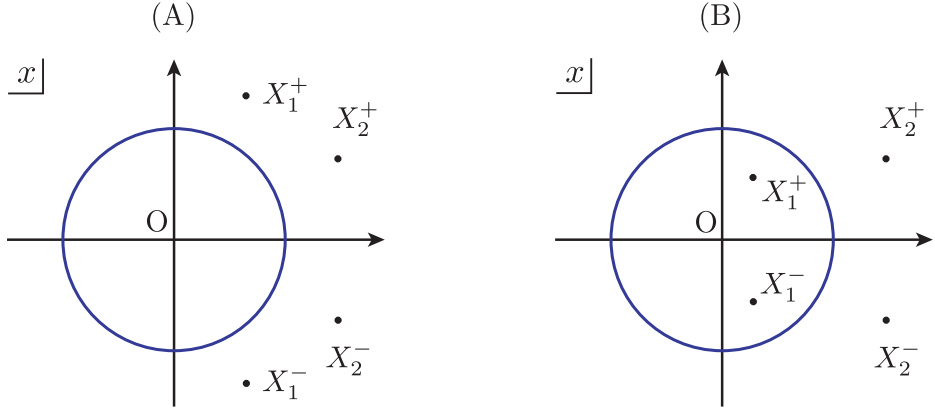}
\vspace{.0cm}
\caption{\small Locations of spectral parameters for Case (A) and (B).
When $X^{\pm}$ are outside the unit circle, the particle has positive charge, otherwise negative.}
\label{fig:spec}
\end{center}
\end{figure}

In Figure \ref{fig:rap} (A-1), the incoming particles are described by $\mathcal C_{1}(Y_{1}^{\pm})\cup \mathcal C_{2}(Y_{2}^{\pm})$\,, where
\begin{align}
{\mathcal C_{1}}(Y_{1}^{\pm};Q_{1})&=\Big\{u_{\tilde\jmath_{1}}\, \Big|\, u_{\tilde\jmath_{1}}-u_{\tilde\jmath_{1}+1}=i/g\,,
\,\,\, \tilde\jmath_{1}=\komoji{Q_{2}-m+1\,, \dots\,, Q_{1}+Q_{2}-m-1}\Big\}\,, \\
{\mathcal C_{2}}(Y_{2}^{\pm};Q_{2})&=\Big\{u_{\tilde\jmath_{2}}\, \Big|\, u_{\tilde\jmath_{2}}-u_{\tilde\jmath_{2}+1}=i/g\,,
\,\,\, \tilde\jmath_{2}=\komoji{1\,, \dots\,, Q_{2}-1}\Big\}\,.
\end{align}
Each cross $(\times)$ represents a pole of the BDS S-matrix.
There is an overlap of length $m-1$ units (one unit is of length $i/g$) running from $u_{Q_{2}-m}$ to $u_{Q_{2}-1}$\,.
One can view this configuration as physically equivalent to $\mathcal M(X_{2}^{\pm})\cup \mathcal N(X_{1}^{\pm})$ of Figure \ref{fig:rap} (A-2), where
\begin{align}
&{\mathcal M}(X_{2}^{\pm};Q_{1}+Q_{2}-m)=\Big\{u_{j_{1}}\, \Big|\, u_{j_{1}}-u_{j_{1}+1}=i/g\,,
\,\,\, j_{1}=\komoji{1\,, \dots\,, Q_{1}+Q_{2}-m-1}\Big\}\,, \\
&\mathcal N(X_{1}^{\pm};m)=\Big\{u_{j_{2}}\, \Big|\, u_{j_{2}}-u_{j_{2}+1}=i/g\,,
\,\,\, j_{2}=\komoji{Q_{2}-m+1\,, \dots\,, Q_{2}-1}\Big\} \,.
\end{align}
They correspond to the intermediate BPS particles, both carrying positive charges.
The locations of the spectral parameters $X_{1,2}^{\pm}$ are shown in Figure \ref{fig:spec} (A).
%
%
They can be expressed by the momenta and charges as
\begin{alignat}{3}
&X_{1}^{+}(P_{1},Q_{1})\eq x_{Q_{2}-m}^{+}=R_{1}\,e^{iP_{1}/2}\,,&\qquad &{}
X_{1}^{-}(P_{1},Q_{1})\eq x_{Q_{2}}^{-}=R_{1}\,e^{-iP_{1}/2}\,;\label{X1}\\
&X_{2}^{+}(P_{2},Q_{2})\eq x_{1}^{+}=R_{2}\,e^{iP_{2}/2}\,,&\qquad &{}
X_{2}^{-}(P_{2},Q_{2})\eq x_{Q_{1}+Q_{2}-m}^{-}=R_{2}\,e^{-iP_{2}/2}\,,\label{X2}
\end{alignat}
where $x=x(u)$ as in (\ref{x(u)}), and $R_{j}=R(P_{j}, Q_{j})$ as in (\ref{radius}).
In terms of these parameters, the charge, momentum and energy of $\mathcal M$ are given by $Q_{1}=Q(X_{1}^{\pm})$\,, $P_{1}=P(X_{1}^{\pm})$ and $E_{1}=E(X_{1}^{\pm})$\,, and the similar for ${\mathcal N}$\,.
For $\mathcal C_{1}$ and $\mathcal C_{2}$\,, we assign spectral parameters $Y_{1}^{\pm}$ and $Y_{2}^{\pm}$ defined by
\begin{alignat}{3}
& Y_{1}^{+}(\tilde P_{1},\tilde Q_{1})\eq x_{Q_{2}-m}^{+}\,,&\qquad &{}
Y_{1}^{-}(\tilde P_{2},\tilde Q_{2})\eq x_{Q_{1}+Q_{2}-m}^{-}\,;\label{Y1}\\
& Y_{2}^{+}(\tilde P_{1},\tilde Q_{1})\eq x_{1}^{+}\,,&\qquad &{}
Y_{2}^{-}(\tilde P_{2},\tilde Q_{2})\eq x_{Q_{2}}^{-}\,.\label{Y2}
\end{alignat}
In terms of these parameters, the charge, momentum and energy of $\mathcal C_{j}$ are given by $\tilde Q_{j}=Q(Y_{j}^{\pm})$\,, $\tilde P_{j}=P(Y_{j}^{\pm})$ and $\tilde E_{j}=E(Y_{j}^{\pm})$\,.
From the definitions (\ref{X1})\,-\,(\ref{Y2}), we see the parameters $X_{1,2}^{\pm}$ and $Y_{1,2}^{\pm}$ are related as
\begin{equation}
Y_{1}^{+}=X_{1}^{+}\,,\quad 
Y_{1}^{-}=X_{2}^{-}\,,\quad 
Y_{2}^{+}=X_{2}^{+}\,,\quad 
Y_{2}^{-}=X_{1}^{-}\,.
\label{X<->Y 2}
\end{equation}
Using the relation (\ref{X<->Y 2}), one can easily check
\begin{equation}
Q_{1}+Q_{2}=\tilde Q_{1}+\tilde Q_{2}\,,\qquad 
E_{1}+E_{2}=\tilde E_{1}+\tilde E_{2}\,,\qquad 
P_{1}+P_{2}=\tilde P_{1}+\tilde P_{2}\,.
\label{identity}
\end{equation}
Therefore the assignments of spectral parameters (\ref{X1})\,-\,(\ref{Y2}) are consistent with the condition that $\mathcal M\cup\mathcal N$ and $\mathcal C_{1}\cup\mathcal C_{2}$ are physically the same.
This condition is, of course, the same as (\ref{XYZ1})-(\ref{XYZ2}) obtained by the physical process analysis.

%
%

\subsubsection*{Dressing part\,:~double poles from ``gaps''}

Let us see how the double poles (\ref{u-u:2}) are seen in the rapidity/spectral plane.
Just as in Case (A), one can interpret the same configuration in two different ways.
One is shown in Fugure \ref{fig:rap} (B-1), $\mathcal C'_{1}(Y_{1}^{\pm})\cup \mathcal C'_{2}(Y_{2}^{\pm})$\,, where
\begin{align}
{\mathcal C'_{1}}(Y_{1}^{\pm};Q_{1})&=\Big\{u_{\tilde\jmath_{1}}\, \Big|\, u_{\tilde\jmath_{1}}-u_{\tilde\jmath_{1}+1}=i/g\,,
\,\,\, \tilde\jmath_{1}=\komoji{Q_{2}+m+1\,, \dots\,, Q_{1}+Q_{2}+m-1}\Big\}\,, \\
{\mathcal C'_{2}}(Y_{2}^{\pm};Q_{2})&=\Big\{u_{\tilde\jmath_{2}}\, \Big|\, u_{\tilde\jmath_{2}}-u_{\tilde\jmath_{2}+1}=i/g\,,
\,\,\, \tilde\jmath_{2}=\komoji{1\,, \dots\,, Q_{2}-1}\Big\}\,,
\end{align}
with gap of length $m-1$ units running from $u_{Q_{2}+1}$ to $u_{Q_{2}+m}$\,.
They correspond to external particles in Figure \ref{fig:dp} (B), see the time-slice (B-1) of Figure \ref{fig:dp}.

The other configuration is shown in Fugure \ref{fig:rap} (B-2), $\mathcal M(X_{2}^{\pm})\cup \overline{\mathcal N}(X_{1}^{\pm})$\,, where
\begin{align}
&{\mathcal M}(X_{2}^{\pm};Q_{1}+Q_{2}+m)=\Big\{u_{j_{1}}\, \Big|\, u_{j_{1}}-u_{j_{1}+1}=i/g\,,
\,\,\, j_{1}=\komoji{1\,, \dots\,, Q_{1}+Q_{2}+m-1}\Big\}\,, \\
&\overline{\mathcal N}(X_{1}^{\pm};-m)=\Big\{u_{j_{2}}\, \Big|\, u_{j_{2}}-u_{j_{2}+1}=i/g\,,
\,\,\, j_{2}=\komoji{Q_{2}-1\,, \dots\,, Q_{2}+m-1}\Big\} \,.
\end{align}
The BPS boundstate with positive charge is described by $\mathcal M(X_{2}^{\pm})$\,, and the one with negative charge is $\overline{\mathcal N}(X_{1}^{\pm})$\,.
In Fugure \ref{fig:rap} (B-2), we depict constituent magnons of $\overline{\mathcal N}$ by a white circle $(\,\circ\,)$ to distinguish it from one with positive charge ($\times$).

In this Case (B), we define the spectral parameters for $\mathcal M(X_{2}^{\pm})$ and $\overline{\mathcal N}(X_{1}^{\pm})$ as
\begin{alignat}{3}
&1/X_{1}^{+}(P_{1},Q_{1})\eq x_{Q_{2}}^{-}=R_{1}\,e^{iP_{1}/2}\,,&\qquad &{}
1/X_{1}^{-}(P_{1},Q_{1})\eq x_{Q_{2}+m}^{+}=R_{1}\,e^{-iP_{1}/2}\,;\label{X1b}\\
&X_{2}^{+}(P_{2},Q_{2})\eq x_{1}^{+}=R_{2}\,e^{iP_{2}/2}\,,&\qquad &{}
X_{2}^{-}(P_{2},Q_{2})\eq x_{Q_{1}+Q_{2}+m}^{-}=R_{2}\,e^{-iP_{2}/2}\,.\label{X2b}
\end{alignat}
Their locations are shown in Figure \ref{fig:spec} (B).
Notice $X_{2}^{\pm}$ reside inside the unit circle, reflecting the associated particle carries negative charge.
The spectral parameters for $\mathcal C'_{1}(Y_{1}^{\pm})$ and $\mathcal C'_{2}(Y_{2}^{\pm})$ are defined as
\begin{alignat}{3}
& Y_{1}^{+}(\tilde P_{1},\tilde Q_{1})\eq x_{Q_{2}+m}^{+}\,,&\qquad &{}
Y_{1}^{-}(\tilde P_{2},\tilde Q_{2})\eq x_{Q_{1}+Q_{2}+m}^{-}\,;\label{Y1b}\\
& Y_{2}^{+}(\tilde P_{1},\tilde Q_{1})\eq x_{1}^{+}\,,&\qquad &{}
Y_{2}^{-}(\tilde P_{2},\tilde Q_{2})\eq x_{Q_{2}}^{-}\,.\label{Y2b}
\end{alignat}
With the assignments of spectral parameters (\ref{X1b})\,-\,(\ref{Y2b}), we see the parameters $X_{1,2}^{\pm}$ and $Y_{1,2}^{\pm}$ are now related as, contrast to (\ref{X<->Y 2}) of Case (A),
\begin{equation}
Y_{1}^{+}=1/X_{1}^{-}\,,\quad 
Y_{1}^{-}=X_{2}^{-}\,,\quad 
Y_{2}^{+}=X_{2}^{+}\,,\quad 
Y_{2}^{-}=1/X_{1}^{+}\,.
\end{equation}
This is of course consistent with (\ref{XYZ1}) and (\ref{XYZ22}).
Using this relation, one can again verify the same conservation conditions as (\ref{identity}).
In summary, the infinitely many possible lengths of the gap between $\mathcal C'_{1}$ and $\mathcal C'_{2}$ (or in other words, the number of roots in $\overline{\mathcal N}$) correspond to infinitely many double poles (\ref{u-u:2}) in the BES phase.

\subsubsection*{Dispersion relation}

Let us see how the dispersion relations for the configurations $\mathcal M\cup\mathcal N$${}={}$$\mathcal C_{1}\cup \mathcal C_{2}$ (Case (A)) and $\mathcal M\cup\overline{\mathcal N}$${}={}$$\mathcal C'_{1}\cup \mathcal C'_{2}$ (Case (B)) look like, in terms of $Q_{j}$\,, $P_{j}$ and $R_{j}$\,.
In both cases, the the total charge and energy are given by the same expressions,
\begin{align}
Q_{1}+Q_{2}&=2g\kko{\Big(R_{1}-\f{1}{R_{1}}\Big)\sin\Big(\f{P_{1}}{2}\Big)+\Big(R_{2}-\f{1}{R_{2}}\Big)\sin\Big(\f{P_{2}}{2}\Big)}\,,\\
E	&=2g\kko{\Big(R_{1}+\f{1}{R_{1}}\Big)\sin\Big(\f{P_{1}}{2}\Big)+\Big(R_{2}+\f{1}{R_{2}}\Big)\sin\Big(\f{P_{2}}{2}\Big)}\,,
\end{align}
and the dispersion relation becomes
\begin{align}
E=
	\sqrt{\ko{Q_{1}+Q_{2}}^{2}+16g^{2}
	\ko{\sin\Big(\f{P_{1}}{2}\Big)+\rho\sin\Big(\f{P_{2}}{2}\Big)}
	\ko{\sin\Big(\f{P_{1}}{2}\Big)+\f{1}{\rho}\sin\Big(\f{P_{2}}{2}\Big)}}\,,
\label{dispersion}
\end{align}
where we defined $\rho\eq R_{1}/R_{2}$\,.
A special case $P_{1}=P_{2}$ reproduces the result obtained in \cite{Spradlin:2006wk} (as a ``boundstate'' of two dyonic giant magnons) after setting $\rho=e^{q}$\,.
Notice that (\ref{dispersion}) can be also expressed as a sum of two BPS particles with positive energies,
\begin{equation}
E=\sqrt{Q_{1}^{2}+16g^{2}\sin\Big(\f{P_{1}}{2}\Big)}+\sqrt{Q_{2}^{2}+16g^{2}\sin\Big(\f{P_{2}}{2}\Big)}\,.
\end{equation}

\paragraph{}
Concerning Case (B), the analyses made in this section gives a support to the observation made in \cite{Dorey:2007xn} from a `quantised' point of view in the following sense.
If we only work in $\mathbb R\times S^{2}$ sector of the theory, the string solution obtained from a breather solution of sine-Gordon equation might seem like a non-BPS boundstate (as was indeed the case when they first appeared in \cite{Hofman:2006xt}), which is absent from the BPS spectrum.
However, it was correctly understood in \cite{Dorey:2007xn} that the ``breathing'' solutions can be and should be interpreted as, once embedded into a larger subspace $\mathbb R\times S^{3}$\,, a superposition of two BPS boundstates with opposite signs for $J_{2}$\,-charge.
This is indeed the picture we have obtained for Case (B);
If we only work in $SU(2)$ sector, the configuration $\mathcal C'_{1}\cup \mathcal C'_{2}$ (Figure \ref{fig:rap} [B-1]) corresponds to a non-BPS state which we cannot find in the BPS spectrum (\ref{E2}), but once we enlarge the sector from $SU(2)$ to $SU(2)\times \overline{SU(2)}$ (which is the same symmetry as the isometry of the $S^{3}$ of string theory), one can view it as a superposition of two BPS boundstates, {\em i.e.}, $\mathcal M$ and $\overline{\mathcal N}$\,, each of which living in a different $SU(2)$ sector (Figure \ref{fig:rap} [B-2]).

\section{Summary and Discussion}

\begin{figure}[tb]
\begin{center}
\vspace{.5cm}
\hspace{-.0cm}\includegraphics[scale=0.85]{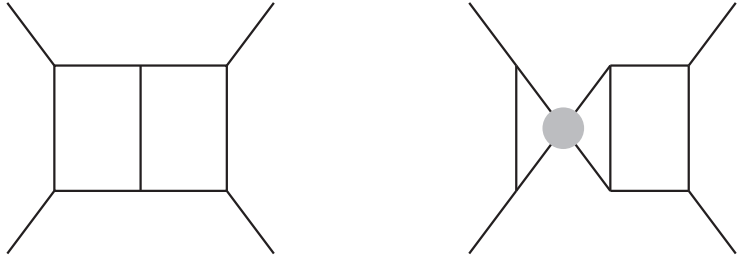}
\vspace{.0cm}
\caption{\small Examples of Landau diagrams that give rise to triple poles.}
\label{fig:tp}
\end{center}
\end{figure}

In this paper we have investigated the singularities of the
bound-state S-matrix which lie near the physical region of real,
postive energy and found a physical explanation for each of them in
terms of on-shell intermediate states. This
is further evidence in favour of the conjectured spectrum and S-matrix
of the ${\cal N}=4$ SYM spin-chain but is by no means a conclusive 
test. There are several ways in which our analysis could be made more
comprehensive. First it would be interesting to extend our analysis to 
triple or higher-order poles. The conjectured S-matrix does not appear
to have such singularities. However, one certainly can draw Landau
diagrams which seem to correspond to triple poles, 
and some of them are shown in Figure \ref{fig:tp}. For consistency 
each of these diagrams must represent an unphysical process for some
reason or cancel in some other way but this remains to be checked. 

Although, we have succeeded in determining the locations of 
physical poles for general cases, we have not determined their residues. As in a
relativistic theory, there may be additional constraints on these
residues for physical intermediate states. Finally, it would be nice
to confirm some of the singularity structure we have described here by
explicit calculations either in gauge theory or on the string
worldsheet.

\subsubsection*{Acknowledgments}
The authors would like to thank Juan Maldacena for comments on the manuscript.
KO is grateful to University of Cambridge, Centre for Mathematical Sciences, for its warm hospitality, during the work was done.
The work of KO is supported in part by JSPS Research Fellowships for Young Scientists.


\appendix

\section{Breathing Magnons\label{sec:BM}}

The string $O(4)$ sigma model with Virasoro constraints is classically equivalent to Complex sine-Gordon (CsG) model, and it is often useful to exploit this connection with CsG theory to construct classical strings.
Indeed, the Pohlmeyer's reduction procedure has been efficiently utilized in the construction of various string solutions, see {\em e.g.}, \cite{PR}.
Our aim here is to realise the oscillating solutions of \cite{Hofman:2006xt, Spradlin:2006wk} from the standpoint of (C)sG solitons.
They correspond to Case (B) of the box diagram discussed in the main text.
In particular, we will concentrate on the $|Q_{1}-Q_{2}|=2$ case.

\subsection{``Minimal'' oscillating solutions from CsG solitons\,...}

The Complex sine-Gordon equation is given by
\begin{equation}
\pa_{+}\pa_{-}\psi +\psi^{*}\,\f{\pa_{+}\psi\pa_{-}\psi}{1-|\psi|^{2}}+\psi (1-|\psi|^{2})=0\,,
\end{equation}
where $\psi(t,x)$ is a complex field and $\pa_{\pm}=\pa_{t}\pm\pa_{x}$ with $(t,x)$ rescaled worldsheet variables $-\infty<t<\infty$ and $-\infty<x<\infty$\,.
It has kink soliton solutions of the form
\begin{equation}
\psi_{\rm K}(t,x)= \f{(\cos\al)\exp\kko{i(\sin\al)\ko{\cosh\Th\cdot t-\sinh\Th\cdot x}}}{\cosh\kko{(\cos\al)\ko{\cosh\Th\cdot x-\sinh\Th\cdot t}}}\,.
\end{equation}
They can be mapped to dyonic giant magnons via the Pohlmeyer's reduction procedure, and they reproduce the dispersion relation for magnon boundstates under identifications 
\begin{align}
E=\f{4g\cos\al\cosh\Theta}{\cos^{2}\al+\sinh^{2}\Theta}\,,\qquad 
Q=\f{4g\cos\al\sin\al}{\cos^{2}\al+\sinh^{2}\Theta}\,.
\label{E,Q}
\end{align}
On string theory side, they represent the energies and the second spins of dyonic giant magnons \cite{Chen:2006ge}, while on gauge theory side, they represent $\Delta-J_{1}$ and the number of constituent $SU(2)$ magnons in the boundstate, respectively.
The spectral parameters of the $Q$\,-magnon boundstates are expressed in terms of the CsG parameters as \cite{Chen:2006gq}
\begin{equation}
X_{j}^{\pm}=\coth\kko{\f{\Th_{j}}{2}\pm i\ko{\f{\al_{j}}{2}-\f{\pi}{4}}}
=\f{\sinh\Th_{j}\pm i\cos\al_{j}}{\cosh\Th_{j}-\sin\al_{j}}\,.
\label{Lambda}
\end{equation}
Recall the parametrisation $X_{j}^{\pm}=R_{j}\,e^{iP_{j}/2}$ introduces before, then the above dictionary tells
\begin{equation}
R_{j}=\sqrt{\f{\cosh\Th_{j}+\sin\al_{j}}{\cosh\Th_{j}-\sin\al_{j}}}\,,\qquad 
\cot\ko{\f{P_{j}}{2}}=\f{\sinh\Th_{j}}{\cos\al_{j}}\,.
\label{R,P}
\end{equation}
Since we are interested in (C)sG description of the oscillating solutions, which are made up of two magnon boundstates with opposite charges, in view of (\ref{E,Q}) we should start with two CsG kinks $j=1,2$ having opposite signs for rotational parameters $\al_{j}$\,.
For simplicity, we restrict our analysis to the ``minimal'' case, $|Q_{1}-Q_{2}|=2$\,, where the length of 
corresponding boundstates $\mathcal M$ and $\overline{\mathcal N}$ differ only by two units.
In this case, in the first approximation in large-$g$\,, the two rotation parameters add up to zero, $\al_{1}=-\al_{2}$\,, and this condition implies the relation between radii as $R_{1}=1/R_{2}$ due to (\ref{R,P}).
Combining this with the condition that the string-centers of $\mathcal M$ and $\overline{\mathcal N}$ coincide, which reads in terms of CsG parameters
\begin{equation}
\f{\sinh(2\Th_{1})}{\cosh(2\Th_{1})+\cos(2\al_{1})}
=\f{\sinh(2\Th_{2})}{\cosh(2\Th_{2})+\cos(2\al_{2})}\,,
\label{U=U' in CsG}
\end{equation}
it follows that $\tanh\Th_{1}=\tanh\Th_{2}$ and $P_{1}=-P_{2}$\,.
In the CsG context, this means two kinks are moving in the same direction with the same velocity.
For notational simplicity, for this ``minimal'' case, we will here set as $R\eq R_{1}=1/R_{2}$\,, $P\eq P_{1}=-P_{2}$\,, $\al\eq \al_{1}=-\al_{2}$ and $\Th_{0}\eq \Th_{1}=\Th_{2}$\,.
The dispersion relation for this solution is then given by
\begin{equation}
E=4g\ko{R+\f{1}{R}}\sin\ko{\f{P}{2}}
\quad \mbox{with}\quad 
R=\sqrt{\f{\cosh\Th_{0}+\sin\al}{\cosh\Th_{0}-\sin\al}}\,,\quad 
\cot\ko{\f{P}{2}}=\f{\sinh\Th_{0}}{\cos\al}\,.
\label{special BM}
\end{equation}
This is the dispersion relation for the minimal oscillating string.
If we set $R=e^{q/2}$\,, it agrees with the dispersion relation obtained in \cite{Spradlin:2006wk} by the dressing method. 

It will become clear in the next section that under proper identification of parameters, the sine-Gordon breathers can be identified with the $\al_{1}=-\al_{2}$ case of CsG kink-kink scattering solutions we have examined, thus also corresponding to the same oscillating string with dispersion relation (\ref{special BM}).

\subsection{...\,and from sG breathers}

\begin{figure}[tb]
\begin{center}
\vspace{.5cm}
\hspace{-.0cm}\includegraphics[scale=0.76]{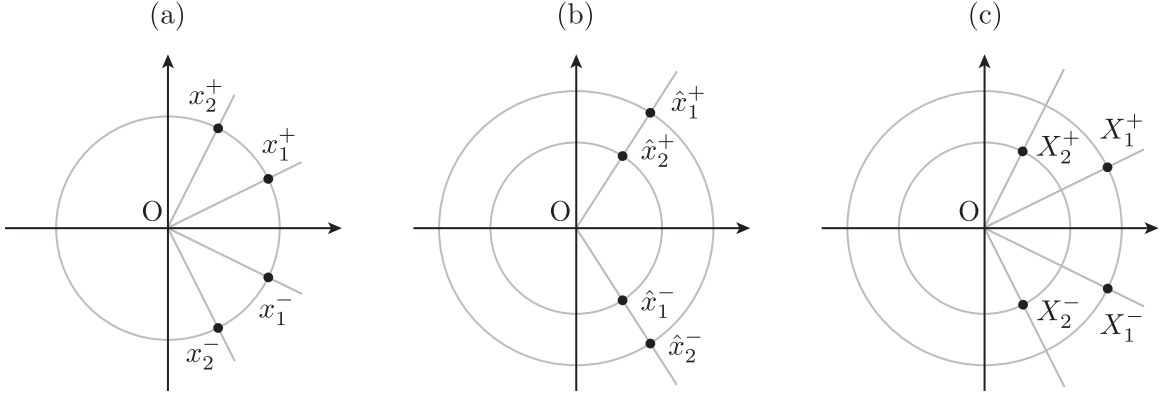}
\vspace{.0cm}
\caption{\small (a): A generic kink-antikink scattering solution of sine-Gordon equation.
(b): A breather solution.
It also describes a special case of CsG kink-kink scattering solution $\al_{1}=-\al_{2}$\,.
(c): A generic CsG kink-kink scattering solution.}
\label{fig:rapidity}
\end{center}
\end{figure}

Let us now turn to sine-Gordon (not ``Complex'') theory to see how this special case of the minimal oscillating string emerges from the sine-Gordon point of view.
The classical sine-Gordon equation,
\begin{equation}
\pa_{+}\pa_{-}\phi-\sin\phi=0\,,
\label{sG}
\end{equation}
has two types of finite-energy solutions.
One is a soliton, which is time-independent and topologically non-trivial solution.
The other is a breather, which is time-dependent and topologically trivial solution, and it can be viewed as a boundstate of a kink and an antikink oscillating in and out, namely, breathing.

\paragraph{}
Let us begin with sG kink-antikink scattering solution.
It is given by
\begin{equation}
\phi_{\rm K\overline{K}}(t,x)=2\arctan\kko{
\f{1}{\tanh\theta}
\f{\sinh\ko{\sinh\theta\cosh\theta_{0}\ko{t-\tanh\theta_{0}\cdot x}}}{\cosh\ko{\cosh\theta\cosh\theta_{0}\ko{x-\tanh\theta_{0}\cdot t}}}
}\,,
\label{kink-antikink}
\end{equation}
where the kink has velocity $\tanh\ko{\theta_{0}+\theta}$ and the antikink has $\tanh\ko{\theta_{0}-\theta}$\,.
Here $\tanh\theta_{0}$ is the velocity of the center-of-mass and $\tanh\theta$ is the relative velocity.
It is convenient to introduce complex spectral parameters
\begin{equation}
x_{j}^{+}=\f{e^{\th_{j}}+i}{e^{\th_{j}}-i}\,,\qquad
x_{j}^{-}=\f{e^{\th_{j}}-i}{e^{\th_{j}}+i}
\end{equation}
with $\th_{1}=\th_{0}+\th$ and $\th_{2}=\th_{0}-\th$\,.
The parameters $x_{j}^{\pm}$ are located on a unit circle in the complex plane and satisfy reality conditions $x_{j}^{+}=(x_{j}^{-})^{*}$\,, see Figure \ref{fig:rapidity} (a).
We also introduce another parametrisation $x_{j}^{\pm}=e^{\pm ip_{j}/2}$\,.
This way of parametrisation is useful in discussing corresponding string solution and also its gage theory dual.
In view of classical-string/sG dictionary, in the limit $t\to \pm\infty$, the profile (\ref{kink-antikink}) corresponds to two giant magnons having angular differences $p_{1}$ and $p_{2}$ between their endpoints.
They in turn correspond to two isolated magnons in an asymptotic SYM spin-chain, each of which having quasi-momenta $p_{1}$ and $p_{2}$\,, respectively.
Note also the relation between $p_{j}$ and $\th_{j}$ are the same as that of $\al_{j}\to 0$ limit of CsG case, see (\ref{R,P}).

\paragraph{}
The sG breather solution can be obtained as an analytic continuation of a kink-antikink scattering solution (\ref{kink-antikink}).
By setting $\theta= i\hat\theta$ in (\ref{kink-antikink}), we obtain
\begin{equation}
\phi_{\rm B}(t,x)=2\arctan\kko{
\f{1}{\tan\hat\theta}
\f{\sin\ko{\sin\hat\theta\cosh\theta_{0}\ko{t-\tanh\theta_{0}\cdot x}}}{\cosh\ko{\cos\hat\theta\cosh\theta_{0}\ko{x-\tanh\theta_{0}\cdot t}}}}\,.
\label{breather}
\end{equation}
The solution (\ref{breather}) represents a breather that is moving with velocity $\tanh\th_{0}$ and oscillating with frequency $f=\sin\hat\th/(2\pi \cosh\th_{0})$\,.
The kink and antikink have complex conjugate velocities $\hat v_{1}=\tanh(\th_{0}+\th)$ and $\hat v_{2}=\tanh(\th_{0}-\th)$\,.
It is again convenient to introduce parametrisations $\hat v_{1}=\cos(\hat p_{1}/2)$ and $\hat v_{2}=\cos(\hat p_{2}/2)$ with $\hat p_{1}=p-iq$ and $\hat p_{2}=p+iq$\,.
These two ways of parametrising the velocities are related through the relations
\begin{equation}
\tan\ko{\f{p}{2}}=\f{\cos\hat\th}{\sinh\th_{0}}\,,\qquad 
\tanh\ko{\f{q}{2}}=\f{\sin\hat\th}{\cosh\th_{0}}\,.
\label{p,q}
\end{equation}
In the kink-antikink scattering case, the rapidities $x_{1}^{\pm}$ (for the kink) and $x_{2}^{\pm}$ (for the antikink) satisfy $x_{j}^{+}=(x_{j}^{-})^{*}$\,, which means both the kink and antikink are physical particles.
In the current breather case, after the analytic continuation, the rapidities become
\begin{align}
\hat x_{1}^{\pm}\eq e^{\pm i\hat p_{1}/2}&=e^{\pm q/2} e^{\pm ip/2}=\f{e^{\th_{0}} e^{i\hat\th}\pm i}{e^{\th_{0}} e^{i\hat\th}\mp i}\,,\\[2mm]
\hat x_{2}^{\pm}\eq e^{\pm i\hat p_{2}/2}&=e^{\mp q/2} e^{\pm ip/2}=\f{e^{\th_{0}} e^{-i\hat\th}\pm i}{e^{\th_{0}} e^{-i\hat\th}\mp i}\,.
\end{align}
This is shown in Figure \ref{fig:rapidity} (b).
In this case we have $\hat x_{1}^{+}=(\hat x_{2}^{-})^{*}$ and $\hat x_{1}^{-}=(\hat x_{2}^{+})^{*}$\,, which means particle $1$ and $2$ are no longer physical particles but instead $\hat x_{1}^{+}\mbox{\,-\,}\hat x_{2}^{-}$ pair (which we call particle $1'$) and $\hat x_{1}^{-}\mbox{\,-\,}\hat x_{2}^{+}$ pair (particle $2'$) represent physical particles.
Actually, these particles $1'$ and $2'$ can be identified with the CsG kinks with $\al_{1}=-\al_{2}$\,, so that they form a minimal breathing solution in the string side.
Explicitly, all physical constraints turn out to be identical under identification $\th_{0}\eq\Th_{0}$ and $\hat\th\eq\al$\,.
If we define $\hat r_{1'}=|\hat x_{1}^{+}|=|\hat x_{2}^{-}|=e^{q/2}$ and $\hat r_{2'}=|\hat x_{1}^{-}|=|\hat x_{2}^{+}|=e^{-q/2}$\,, then one can also check $R_{1}=\hat r_{1'}$ and $R_{2}=\hat r_{2'}$ in this case.

%

\paragraph{}
In view of the second equality in (\ref{p,q}), the period of the oscillation is also expressed as $T=1/f=2\pi/\tanh(q/2)$\,.
This is different from the period of rotation by the factor of $\tanh(q/2)$\,; these two kinds of periods agree only in the limit $q\to \infty$\,.
As a result, while the location of endpoints of a string on the equator are the same between $t=0$ and $t=T$\,, the shape of the string itself are not in general.

As was done in \cite{Hofman:2006xt}, one can relate the parameter $q$ which controls the period of breathing to the oscillation number $n$ of breather solution.
The energy of elementary magnon is given by the formula $\ep_{j}=4g\sin(\hat p_{j}/2)$\,, which is the large-$g$ limit of (\ref{Delta}) (or large-$g$ limit of (\ref{E}) with $Q=1$).
Each constituent magnon has complex energy, but since they are complex conjugate to each other, they sum up to give a real energy for the minimal breathing solution, 
$\ep_{\rm BM}=\ep_{1}+\ep_{2}$\,.
We can then define the oscillation number as the action variable associated with the breathing,
\begin{equation}
n=\int \f{T}{2\pi}\,d\ep_{\rm BM}
=8g \sin\ko{\f{p}{2}}\sinh\ko{\f{q}{2}}\,.
\end{equation}
Large $n$ thus means large value of $q$ for fixed $p$\,.
As noted before, in the limit $n\to \infty$ or $q\to \infty$\,, the two kinds of periods, the oscillation period and the rotation period become identical (both are equal to $2\pi$).
This means that when $p$ is near to $\pi$\,, the string looks like no more rotating but rather pulsating; staring from one point on the equator and sweeps whole the sphere and shrinks into its antipodal points, and again back to the original point by time reversal motion with only the orientation changed.



\providecommand{\href}[2]{#2}\begingroup\raggedright\endgroup

\end{document}